\documentclass[12pt,preprint]{aastex}

\usepackage{emulateapj5}

\newcommand{\et}{et al.}

\newcommand{\Dtbin}{{\Delta}T_{{\rm bin}}}
\newcommand{\Dtsim}{{\Delta}T_{{\rm sim}}}
\newcommand{\fmin}{f_{{\rm min}}}
\newcommand{\fNyq}{f_{{\rm Nyq}}}

\newcommand{\xsqdist}{\chi ^2 _{{\rm dist}}}

\newcommand{\Dtsamp}{{\Delta}T_{{\rm samp}}}
\newcommand{\rxte}{{\it RXTE}}
\newcommand{\xte}{{\it RXTE}}

\newcommand{\xmm}{{\it XMM-Newton}}
\newcommand{\chandra}{{\it Chandra}}

\newcommand{\Msun}{\hbox{$\rm\thinspace M_{\odot}$}}

\begin{document}
\title{X-ray Fluctuation Power Spectral Densities of Seyfert~1 Galaxies}

\author{A.~Markowitz\altaffilmark{1}, R.~Edelson\altaffilmark{1},
S.~Vaughan\altaffilmark{2}, P.~Uttley\altaffilmark{3},
I.~M.~George\altaffilmark{4,5}, R.~E.~Griffiths\altaffilmark{6},
S.~Kaspi\altaffilmark{7},
A.~Lawrence\altaffilmark{8},
I.~McHardy\altaffilmark{3},
K.~Nandra\altaffilmark{5,9},
K.~Pounds\altaffilmark{10},
J.~Reeves\altaffilmark{10},
N.~Schurch\altaffilmark{10},
R.~Warwick\altaffilmark{10}}
\altaffiltext{1}{Dept.\ of Astronomy, Univ.\ of California, Los Angeles CA
90095-1562; agm@astro.ucla.edu, rae@astro.ucla.edu}
\altaffiltext{2}{Institute of Astronomy, Madingley Road, Cambridge, CB3~0HA, UK; sav@ast.cam.ac.uk}
\altaffiltext{3}{Dept.\ of Physics and Astronomy, University of
Southampton,
Southampton, SO17~1BJ, UK; pu@astro.soton.ac.uk}
\altaffiltext{4}{Joint Center for Astrophysics, Physics Department, University of Maryland, Baltimore County, 1000 Hilltop Circle, Baltimore, MD 21250}
\altaffiltext{5}{Laboratory for High Energy Astrophysics, NASA/Goddard Space Flight Center, Code 662, Greenbelt, MD 20771}
\altaffiltext{6} {Dept.\ of Physics, Carnegie Mellon University, Pittsburgh PA 15213-3890 }
\altaffiltext{7}{School of Physics and Astronomy and the Wise Observatory,
The Raymond and Beverly Sackler Faculty of Exact Sciences, Tel Aviv
University, Tel Aviv 69978, Israel}
\altaffiltext{8}{Institute for Astronomy, University of Edinburgh, Royal Observatory, Blackford Hill, Edinburgh,
EH9~3HJ,~UK}
\altaffiltext{9}{Astrophysics Group, Imperial College, Blackett Laboratory, Prince Consort Road, London, SW7~2BW,~UK} 
\altaffiltext{10}{X-ray Astronomy Group, Univ.\ of Leicester, Leicester,
LE1~7RH,~UK}

\begin{abstract}

By combining complementary monitoring observations spanning long, medium
and short time scales, we have constructed power spectral densities (PSDs)
of six Seyfert~1 galaxies.
These PSDs span $\gtrsim$4 orders of magnitude in temporal
frequency, sampling variations on time scales ranging from tens of minutes
to over a year.
In at least four cases, the PSD shows a "break," a significant departure
from a power law, typically on time scales of order a few days.
This is similar to the behavior of Galactic X-ray binaries (XRBs),
lower mass compact systems with breaks on time scales of seconds.
NGC~3783 shows tentative evidence for a doubly-broken power law, a
feature that until now has only been seen in the (much better-defined)
PSDs of low-state XRBs.
It is also interesting that (when one previously-observed object is added
to make a small sample of seven), an apparently significant correlation is
seen between the break time scale $T$ and the putative black hole mass
$M_{\rm BH}$, while
none is seen between break time scale and luminosity.
The data are consistent with the linear relation
$ T = M_{\rm BH}/10^{6.5} \Msun$; extrapolation over 6--7
orders of magnitude is in reasonable agreement with XRBs.
All of this strengthens the case for a physical similarity between
Seyfert~1s and XRBs.

\end{abstract}


\section{ Introduction }

Early X-ray observations of Seyfert~1 galaxies showed strong, aperiodic
X-ray variability, evidence that the X-rays are emitted in close proximity
to the central black hole.
Because its properties are so well-studied and understood (e.g., Priestley
1981), the fluctuation power spectral density (PSD) is a common tool
for temporal analysis.
The first Seyfert~1 PSDs were measured with the {\it EXOSAT} long-looks.
These established the red-noise nature of Seyfert~1 galaxy variability
over the time scales of $\sim2~\times~10^{-3}$ to $\sim2~\times~10^{-5}$~Hz 
(Lawrence \et\ 1987; McHardy \& Czerny 1987;
Lawrence \& Papadakis 1993; Green, McHardy \& Lehto 1993).

Edelson \& Nandra (1999; hereafter EN99) completed the
first systematic broadband PSD study using a series of
contemporaneous, evenly-sampled {\it Rossi X-ray Timing Explorer} ({\it
RXTE}) observations of
NGC~3516 to measure a PSD spanning over 3 decades of temporal frequency 
($ 4~\times~10^{-8}~-~7~\times~10^{-4}$~Hz).
This yielded the first clear evidence of a break in the power law PSD at
$\sim4~\times~10^{-7}$~Hz ($\sim$1~month) and an intriguing
similarity to the PSDs of X-Ray Binaries (XRBs; 
see also McHardy 1988, Papadakis \& McHardy 1995)

Pounds et al.\ (2001) applied a similar sampling pattern to the
Narrow-Line (Soft-Spectrum) Seyfert~1 Akn~564.
The PSD showed a higher cutoff frequency, $\sim9~\times~10^{-7}$~Hz
(corresponding to $\sim$2~week) which Pounds \et\ (2001)
argued was consistent
with such objects having lower mass black holes emitting at a higher
fraction of the Eddington rate. 
Uttley, McHardy \& Papadakis (2002, hereafter UMP02) confirmed the
break in the PSD of NGC~3516 and additionally studied three other Seyfert
galaxies.  The Seyfert~1 MCG--6-30-15 and the Seyfert~2 NGC~5506 both
showed breaks at $ \sim 5 \times 10^{-5} $~Hz (corresponding to
$\sim$6~hr) but no clear break was seen in the Seyfert~1 NGC~5548.

Given that the above frequency breaks depend on the model used to fit the PSD
as well as the analysis method, a more systematic picture of Seyfert~1
PSDs can be attained with uniform analysis and models.
To accomplish this, we have
applied the EN99 observing technique to a sample of six Seyfert~1 galaxies, as
given in Table~1. 
These monitoring data were used to construct high dynamic range PSDs and
search for departures from power law behavior.
The observations, data reduction and resulting light curves are described
in \S~2.
The PSD measurement and modeling is discussed in \S~3.
We employ the Monte Carlo technique of UMP02
to quantify errors and account for red-noise leak and
aliasing.
The model fitting results and evidence for a break in the power law PSD
are given in \S~4.
The results were then combined with the MCG--6-30-15 result of 
UMP02 to create a PSD survey of seven Seyfert~1 galaxies.
The physical implications of this break is discussed and a comparison with
the PSD and other variability properties of XRBs is made in \S~5.
A brief summary of these results is given in $\S$~6.

\section{Observations and data reduction}

\subsection{Sampling}

The sampling pattern employed herein (based on EN99) covers complementary
long, medium and short time scales with progressively shorter and denser
relatively even sampling.
This makes it possible to construct a PSD that covers the maximum temporal
frequency range while minimizing the amount of telescope time needed.
It also requires the assumption that the variations are stationary or at
most mildly non-stationary; this assumption is tested in \S~3.4.
The specific sampling patterns are described below and tabulated in
Table~2.

\begin{itemize}

\item {\bf Long (months-year) time scales:}
For all six targets, \xte\ obtained nearly even sampling every $\sim$4.3~d
($\sim$64 orbits) over a period of $\gtrsim$3 years.
In NGC~5548, observations in 1996--1999 every $\sim$14~d were also
included.
Observations lasted only a fraction of an orbit ($\sim$1~ksec) in all cases.

\item {\bf Medium (day--weeks) time scales:}
For four of the objects, medium time scale sampling was obtained with \xte\
by observing once every other orbit (3.2~hrs) 
for 256 epochs, spanning 34 days.
For NGC~3783, this sampling was once every other orbit for 151 epochs
($\sim$20 days), and for NGC~3516, it was once every eight orbits for 136
days. Again, $\sim$1~ksec samples were used.

\item {\bf Short (hours) time scales:}
Quasi-continuous ($\sim$60--300 ksec) observations were obtained with
\xmm\
(for NGC~5548 and NGC~4151), \chandra\ (NGC~3783) and \xte\ (NGC~3516).
These observations were interrupted only by periods of increased particle
backgrounds (for \xte\ and occasionally \xmm) and were sensitive to
$\gtrsim$3\% variations on time scales as short as $\sim$1~ksec.
The other two sources (Akn~564 and Fairall~9) had only short ($<$35~ksec)
uninterrupted \xmm\ observations.

\end{itemize}

\subsection{{\it RXTE} data reduction}

All of the long- and medium-term data were obtained with
\xte\ using similar sampling patterns; data reduction proceeded in a
similar fashion for all
these observations, as well as the two NGC~3516 long-looks.
These data were obtained with the
proportional counter array (PCA), which consist of five identical
collimated
proportional counter units (PCUs; Swank 1998).
For simplicity, data were collected only from those PCUs
which did not suffer from repeated breakdown during on-source time
(PCUs 0, 1, and 2 prior to 1998 December 23; PCUs 0 and 2 from 
1998~December~23 until~2000~May~12; PCU~2 only after 2000 May 12).
Count rates quoted in this paper are normalized to 1 PCU.
Only PCA STANDARD-2 data were considered.
The data were
reduced using standard extraction methods and {\sc FTOOLS~v4.2}
software. Data were rejected if they were
gathered less than 10$\arcdeg$ from the Earth's limb, if
they were obtained within 30~min after the satellite's passage
through the SAA, if {\sc ELECTRON0}~$>$~0.1
({\sc ELECTRON2} after 2000~May~12),
or if the satellite's pointing offset was greater
than 0$\fdg$02.

As the PCA has no simultaneous background monitoring capability,
background data were estimated
by using {\sc pcabackest~v2.1b} to generate model files
based on the particle-induced
background, SAA activity, and the diffuse X-ray background.
This background subtraction is the
dominant source of systematic error in \xte\ AGN monitoring data
(e.g., EN99).
Counts were extracted only from the topmost PCU layer to maximize
the signal-to-noise ratio.
All of the targets were faint ($<$~40~ct~s$^{-1}$~PCU$^{-1}$),
so the applicable 'L7-240' background models were used.
Because the PCU gain settings changed
three times since launch, the count rates were rescaled to a common
gain epoch (gain epoch 3) by calibrating with several public archive
Cas~A observations.
Light curves binned to 16~s were generated for all targets over the
2--10~keV bandpass, where the PCA is most sensitive and the
systematic errors and background are best quantified.
The data were then further binned as listed in Table~2, column~5;
bins with less than
10 flux points were excluded from analysis. Standard errors were
derived from the data in each orbital bin.
Further details of \xte\ data reduction can
be found in EN99.

\subsection{{\it XMM-Newton} observations and data reduction}

\subsubsection{NGC~5548}

{\it XMM-Newton} observed NGC~5548 for 97~ksec during revolution 290
from 2001~Jul~09 16:08:04~UT to 2001~Jul~10 18:06:21~UT.
Data from the high-throughput European Photon Imaging Camera
(EPIC) instruments, the pn (Struder \et\ 2001) and two
MOS cameras (Turner \et\ 2001) were used.
The medium filters were used.
Due to their higher count rate and signal-to-noise, only
the pn data were used for PSD analysis; the MOS data are henceforth
ignored.
The pn camera was operated in Small Window mode;
the readout time was 5.7~ms.

A standard reduction of the raw pn data was done using
the Science Analysis System. This involved the subtraction of
hot, dead, or flickering pixels, removal of events due to electronic noise
and correction of event energies for charge transfer losses.
The patterns used were 0--4 (single and double pixel events).
The extraction region used was a circle of radius 40$\arcsec$.
The core was not excluded, as the level of pile-up was
fairly small ($<$2$\%$).
The background count rate was extracted in 250 second bins
using  the same size region in the same chip, but off-source
($>$1$\arcmin$).
Solar flare activity caused a tremendous increase
in the soft proton flux during the final 13.5 ksec, rendering that
section of data useless for PSD analysis.
This resulted in 83.5~ksec of useful data; a 
background-subtracted light curve over the 2--10~keV band
was initially extracted in 1~s
bins. In a red-noise PSD derived from a continous
light curve, the power amplitude
at the highest temporal frequencies will be dominated by the
constant power level contributed by Poisson noise, $P_{\rm Psn}$,
instead of the intrinsic source variability.
To ensure a high variability-to-noise PSD with the intrinsic source
varaibility power being greater than the Poisson noise power at
all frequencies sampled, the data were binned
at the time scale where the intrinsic variability was equal to
$P_{\rm Psn}$; in the case of the NGC~5548 data;
this time scale was 900~s.

\subsubsection{NGC~4151}

NGC~4151 was observed by \xmm\ for 100 ksec during revolution 190
from 2000~Dec~22 02:48:22~UT to 2000~Dec~23 05:29:59~UT.
Due to two 8~ksec gaps which would have seriously complicated the
PSD analysis, only the final 57~ksec was used.
Again, the medium filter was used for all three EPIC instruments,
which were operated in Full Frame mode.
Data reduction over the 2--10~keV band
proceeded in a similar manner to that of NGC~5548;
again, due to its higher count rate and signal-to-noise,
the pn data only were used for PSD analysis.
The patterns used were 0--4 (single and double pixel events).
The area of source extraction was a circle of radius 30$\arcsec$;
background count rates were extracted over a region 4.4 times
as large.
Due to the low source flux, photon pile-up was not an issue.
Background-subtracted light curves initially extracted in 1~s bins
were rebinned to 1100~s.

\subsection{{\it Chandra} observations and data reduction for NGC~3783}

{\it Chandra} observed NGC~3783 for 170~ksec each of five times
during 2001 March--June
using the High Energy Transmission Grating Spectrometer (HETGS; 
e.g., Markert \et\ 1994) with the Advanced CCD Imaging
Spectrometer (ACIS; e.g., Nousek \et\ 1998; Plucinsky \et\ 2002)
as the detector.
The data were reduced using CIAO software version 2.1.2 and its associated
calibration data.
Photons in the 2--10~keV band were extracted from the MEG and HEG dispersed
first orders, as the zeroth order in the HETGS images of NGC~3783 suffers
from heavy photon pile-up.
The extraction width on the dispersed spectrum was
done with the CIAO default of 4.78$\arcsec$.
NGC~3783 is a point like source and no extended circumnuclear
emission was detected;
the working assumption was that all photons in the dispersive arms
were from the central source.
The background count rate was negligible
(several orders of magnitude smaller than the signal) in the dispersed
arms,
and was not subtracted.
Further details of the \chandra\ data reduction can be found in
Kaspi \et\ (2002).
Light curves initially extracted in bins of 32.41~s were
truncated to equal lengths of 167.6 ksec and rebinned to
2000~s for a total of 84 data points in each light curve.

\subsection{The light curves}

The resulting light curves are shown in Figures~1--4.
Figure~1 shows the long- and medium-term data from \rxte\ monitoring,
with the short-term NGC~3516 \xte\ long-looks included and the
locations of the \xmm\ and \chandra\ long-looks marked.
Figure~2 shows the long-term light curves resampled from the
raw \xte\ light curves. Figure~3 shows the resampled
medium-term light curves, and Figure~4 shows the intensive long-looks.
Table~2 summarizes the sampling parameters for each light curve,
including the instrument, mean count rate, sampling interval,
total points, and percent of missing data for
each time scale for which there is even sampling suitable for PSD
analysis.
Also listed is F$_{\rm var}$, the 
square root of the excess variance, a measure of the
intrinsic variability amplitude normalized to the mean
as defined in, e.g., Edelson et~al.~(2002).
We note that $\xte$-PCA, $\chandra$-HETG, and $\xmm$ EPIC pn
do not have precisely identical
responses over the 2--10~keV bandpass; the PCA response is somewhat harder
(peaking closer to $\sim$5~keV) than that of the pn or $\chandra$-HETG 
(which peak closer to $\sim$2~keV). It has been shown that measured
high-frequency PSD slope is energy dependent in both XRBs and Seyferts
(Nowak \et\ 1999a, Nandra \& Papadakis 2001), flattening as photon energy 
increases; our use of \xte\
data at low frequencies and \xmm\ or \chandra\ at high frequencies
could thus potentially mimic a high-frequency break.
However, we expect this
effect to be relatively minor, as PSD slope typically flattens by only 
one or two tenths in power law slope for a doubling or tripling
of photon energy (Nowak \et\ 1999a,
Nandra \& Papadakis 2001), and XRB and PSD breaks typically 
involve a change in power law slope near 1.

\section{PSD construction}

The PSD construction required several steps.
First, PSDs were measured for each individual short, medium or long
time scale light curve and then the individual PSDs were combined to form
one high-dynamic range PSD for each target, as described in \S~3.1.
However, these combined PSDs have poorly determined errors (dominated by
systematic effects such as aliasing and red-noise leak, as described in
\S~3.2).
A Monte Carlo technique was employed to better estimate PSD fit parameters in
the presence of these systematic errors and distortion effects, as
described in \S~3.3.

\subsection{Initial measurement}

First, light curves were linearly interpolated for missing data points,
though such gaps were relatively rare.
Light curves were not interpolated to a strictly even grid,
since departures from ephemeris were almost always
small (typically only a few percent) and such an interpolation
would have negligible impact on the resulting periodogram.
Each light curve's mean was subtracted.

The periodograms were constructed using a Discrete Fourier
Transform (DFT; e.g., Oppenheim \& Shafer 1975; Brillinger 1981).
The power at each Fourier frequency $f~=~1/D, 2/D$, ... $1/(2 \Delta T)$
(the periodogram, where $\Delta T$ is the sampling time) was calculated
using the fractional RMS squared normalization
\[ P(f) = \frac{2D}{\mu^2 N^2} \vert F(f) \vert ^2 \]
where D is the duration, N is the number of points, $\mu$ is the
mean count rate
of the data, and $\vert F(f) \vert ^2$ is the modulus-squared of the
Fourier Transform of the light curve. The null, or rectangular,
lag window was used;
use of a tapered lag window for steep red-noise PSD data 
strongly increases the bias of the periodogram (see e.g., Priestley 1981).
Following 
Papadakis \& Lawrence (1993a), the logarithm of the
periodogram was binned every factor of 1.4 (0.15 in the logarithm)
in temporal frequency, but the lowest frequency bins were widened to
accommodate a minimum of 2 periodogram points.
The constant level of power due to
Poisson noise was not subtracted, but it was modelled in the
Monte Carlo analysis. The individual PSDs were then combined
to form broadband PSDs for each target.

These high-quality broadband PSDs spanned an exceptional
range of time scales: four cover more than 4.4 decades
of temporal frequency and the other two cover 3.6 decades.
PSD measurement parameters, including the minimum and Nyquist
frequencies sampled in each PSD segment, are given in Table~3.
No renormalization in power amplitude of the individual PSDs
was done.
The resulting broadband PSDs are shown in Figure~5.
Visual inspection reveals a variety of behavior; e.g.,
low-frequency flattening is readily evident in the
PSDs of NGC~3783 and Akn~564.

\subsection{Aliasing and red-noise leak}

Although temporal analysis techniques such as the PSD are derived assuming
continuous observations of infinite duration, such conditions cannot be
attained in practice.
These real-life Seyfert~1 sampling patterns $S(t)$ 
span a limited duration
$D$ and have a shortest time resolution $\Dtsamp$.
As the Fourier transform of the observed data is a 
convolution of the Fourier transform of the
underlying variability process with $\hat{S}(f)$,
the 'sampling window,' power is transferred
into the sampled frequency range from above $\fNyq = 1/(2\Dtsamp)$
and below $1/D$ (red-noise leak), as explained below.

If a red-noise PSD contains significant power below $\fmin = 1/D$, then
there may be significant long-term trends present in the light curve.
These trends can dominate the total variance of the observed light curve,
whose measured PSD then contains additional power transferred from below
$\fmin$.
The power transferred by this red-noise leak process has a power-law
slope which goes as $f^{-2}$ (see details in van der Klis 1997).

On short time scales, the discrete sampling causes aliasing.
For bins of width $\Dtbin$ evenly spaced $\Dtsamp$ apart (with 
$\Dtbin~\ll~\Dtsamp$), variations on time scales shorter than $\Dtsamp$ cannot
be distinguished from longer-time scale variations, and power from above
$\fNyq$ is effectively added into the frequency range sampled by the PSD.

Distortion effects are present in, and hamper the interpretation of,
the PSDs constructed above, but another serious problem with these PSDs
is that the lowest temporal frequency bins
contain too few PSD points for normal errors to be assigned.
By binning a sufficient number of periodogram estimates ($\gtrsim$20 for
logarithmically binned periodograms; Papadakis \& Lawrence 1993a), the
averaged power will approach a Gaussian distribution, and the mean of the
periodogram points in a frequency bin will tend to the power amplitude of the
true underlying variability process, $ P_{true} (f) $.

\subsection{Monte Carlo simulations }

To obtain a PSD shape with adequate errors, and to solve the problem of
distortion effects, we use a version of Monte Carlo technique {\sc PSRESP}
first introduced by UMP02, based on a similar Monte Carlo 
technique by Done \et\ (1992).
This enables us to quantify the degree of low-frequency flattening and
find a best-fit model PSD shape corresponding to the underlying
variability process.

The technique consists of simulating light curves from a given PSD model
shape specified for testing, resampling these light curves and measuring
their PSDs in the same manner as the observed light curves, forming an
'average model' broadband PSD which accounts for the distortion effects.
Errors are assigned for all frequency bins,
derived from the RMS spread of the individual simulated PSDs at a given
frequency bin. 
Finally, the technique compares the model broadband PSD to that derived
from the observed data using
a fit statistic with the 
distribution estimated from the simulations to determine 
goodness of fit for the model.
In this manner, a variety of underlying PSD model shapes can be tested
against the data.
We note, as a caveat, that the results of this 
technique are highly model dependent;
specifically, the primary assumption governing this entire process is that
the broadband PSD model shape used for testing is an accurate
representation of the underlying variability process $P_{true}(f)$.
An outline of the Monte Carlo method is as follows
(see UMP02 for further details):

1. An underlying model PSD shape (a continuous function)
is specified for testing. The initial normalization is arbitrary.
For each long-, medium-, and short-term PSD segment, the
algorithm of Timmer \& K\"{o}nig (1995) is used to generate $N$ light curves
(where $N$ is at least 100).
The time resolution $\Dtsim$ is chosen to be 0.1$\Dtsamp$.
To ensure that the light curves account for variability on
time scales longer than the observed light curve
(contain a red-noise leak contribution if necessary), one
very long segment of length $N~\times~D$ is simulated; this is then
broken up into $N$ light curves, each of duration $D$.

2. The light curves are resampled in the same manner as the observed
light curves as follows: For long- and medium-term simulated data, 
every tenth point is selected to
degrade the resolution from $\Dtsim~=~0.1 \Dtsamp $ to
$\Dtsamp$; this correctly accounts for 
most of the total aliasing, specifically that portion of the aliased
power which is due to variations on time scales
from $\fNyq$ to $1/(2\Dtsim)$. 
For continuous long-look simulated data,
the data are averaged over every ten points.
The uncertainty in the
quantity of aliased power in red-noise PSD is dominated 
by the uncertainty in
the amount of power at frequencies just above $\fNyq$, which is why
this resampling procedure must be used.
Time bins with missing flux in the observed light curves are
linearly interpolated.

3. PSDs are constructed from the simulated
light curves in the same manner as the
observed PSDs, using the same
normalization and frequency binning.
For each segment, the model average PSD
$\overline{P_{{\rm sim}}(f)}$ is calculated
from the N individual PSDs; error bars 
$\Delta~\overline{P_{{\rm sim}}(f)}$
equal to the RMS spread of the 
individual PSDs at each frequency bin are assigned.

4. While the resampling approach in step~2 accounts for
much of the total aliasing in red-noise PSDs, there remains aliasing due
to variations on time scales from $\Dtsim$ down to $\Dtbin$.
This second, smaller quantity of
aliased power is approximated
by adding a constant level of power to all frequencies 
sampled in a given PSD.
Since variations on time scales shorter than $\sim2\Dtbin$
will likely not contribute to aliasing, this
quantity of aliased power 
can be approximated by the analytical expression
\[ P_{alias} = \frac{1}{\fNyq - \fmin} \int^{(2\Dtbin)^{-1}}_{(2\Dtsim)^{-1}}
P(f) df \] (e.g., see UMP02 $\S$~3.3)
and added to the model average PSD and the $N$ individual PSDs.
The resulting PSDs then account fully for both red-noise leak
and both aliasing components.

5. The goodness of fit is determined.
Comparison of a model-average PSD to the observed PSD is not possible
in the ``traditional'' $\chi^2$ sense, since, as previously
discussed, error bars
assigned to the observed PSD are not strictly Gaussian.
Instead, a new statistic $\xsqdist$ is defined
to compare the average distorted model PSD to the observed data, using
the well-determined errors from the model:
\[ \xsqdist = \sum_{f} \frac{ ( \overline{P_{{\rm sim}}(f)} -  P_{{\rm
obs}}(f))^2}{ (\Delta \overline{P_{{\rm sim}}(f)})^2} \]
The total $\xsqdist$ value is calculated by summing over all frequency
bins in all PSD segments. For the targets with multiple short-term PSDs
(NGC~3516 and NGC~3783), we average the values of short-term
$\xsqdist$ to avoid unnecessarily
weighting the fit towards the highest frequencies.

The best-fitting normalization of the P$_{{\rm sim}}$(f) is found by
renormalizing all segments of the P$_{{\rm sim}}$(f) by the same factor
until the total $\xsqdist$ value is minimized. The power 
contribution from Poisson
noise is added to the model PSDs during this step;
$P_{\rm Psn}$ is calculated via 
$2(\mu + B)/\mu^2$, where $\mu$ and $B$ are the total source
and background count rates, respectively. For non-continuously observed
light curves, this must be multiplied by $\Dtsamp/\Dtbin$.
The observed-$\xsqdist$ value is calculated. 

The goodness of fit is then determined as follows:
10,000 combinations the of long-, medium-, and short-term simulated PSDs
are
randomly selected to model
the $\xsqdist$ distribution for each given PSD model. For each
combination, the value of simulated-$\xsqdist$ (comparing P$_{{\rm
sim}}(f)$
to $\overline{P_{{\rm sim}}(f)}$) is calculated.
These 10,000 measurements of the simulated-$\xsqdist$ distribution are
sorted into ascending order. The probability that the model PSD
can be rejected is given by the percentile of the
10,000 $\xsqdist$ exceeded by the value of observed $\xsqdist$.

6. The above steps are repeated to test a range of PSD model shapes
by stepping through a range of high-frequency slopes and break
frequencies.  

\section{Results}

The results of the Monte Carlo analysis are presented for
several simple PSD model shapes to quantifying the degree of
flattening, if any, towards low temporal frequencies.
First, unbroken PSD models are tested ($\S$~4.1).
Then, models incorporating a single PSD break are tested
($\S$~4.2).
We do not test for quasi-periodic oscillations 
(QPO); though they are routinely seen in XRB PSDs,
there is no obvious indication of QPO in the observed
light curves or resultant PSDs, and there has been no convincing
evidence to date for deterministic 
behavior in AGN light curves or PSDs
(e.g., Lawrence \et\ 1987, McHardy \& Czerny 1987)
as previous claims of QPO in AGN (e.g., Papadakis \& Lawrence 1993b) 
have all either been refuted or lack confirmation.
Preliminary evidence for a doubly-broken power law model fit to
the PSD of NGC~3783 is presented in $\S$~4.3.
Finally, a test of one of the assumptions governing the PSD construction,
the assumption of PSD stationarity, is presented in \S~4.4.

\subsection{Unbroken power law models}

The simplest model tested was an unbroken power law, of the form $ P(f) =
A_o (f/f_o)^{-\beta}$, where the normalization $A_o$ is the PSD amplitude at $ f
= f_o $, and $\beta$ is the power law slope.
(The constant level of power from Poisson noise is added to each simulated PSD,
but as this value is different for each long-, medium-, and short-term PSD
segment, it is not explicitly listed here.)
For all targets, the model was tested by stepping through the range of
$\beta$ from 1.0 to 4.0 in increments of 0.05; for Akn~564 and NGC~3783,
the two targets with relatively flatter observed PSDs, values of $\beta$
down to 0.0 were additionally tested.
Five hundred simulated PSDs were used to determine the average model PSD,
and 10$^4$ randomly selected sets of PSDs were used to probe the simulated
$\xsqdist$ distribution.

The best-fitting models and simulated data are
plotted in $f~\times~P_f$ space
in Figure~6 for all targets.
The best-fitting values of $\beta$
and $A_o$, along with the 
corresponding likelihood of acceptance (defined as 1 -- $R$, where $R$
is the rejection
confidence)  are summarized in Table~4.
The errors on $\beta$ correspond to 
values 1$\sigma$ above the likelihood
of acceptance for the best-fit value on a Gaussian probability
distribution (i.e., the amount $\beta$ needs to change for the
fit to be less likely by 1$\sigma$).
While we have not yet rigorously
proven the validity of these errors, this method does
give very reasonable values. 
The errors on $A_o$ are determined from the RMS spread of 
the 10$^4$ randomly selected
sets of simulated PSDs.
The low likelihood of acceptance for NGC~3783, for instance, indicates that
an unbroken power law is an inaccurate description of the data.

\subsection{Singly-broken power law models}

To test for the presence of a PSD break, we employed a broken power law
model of the form

\[P(f)= \left\{ \begin{array}{ll}
                              A(f/f_c)^{- \gamma},  & f \le f_c \\
                              A(f/f_c)^{- \beta},   & f > f_c \end{array}
\right. \]
where the normalization $A$ is the PSD 
amplitude at the break frequency $f_{c}$,
$\beta$ is the high frequency power law slope, and
$\gamma$ is the low frequency power law slope, with the constraint 
$\gamma < \beta $.
The range of slopes tested was $\beta=0.0-4.0$ in increments of 0.05.
Break frequencies were tested in the log from --8.0 to --4.0
in increments of 0.1, corresponding to
$f~\rightarrow~1.26f$ in the linear scale.
100 simulated PSDs were used to determine the average model PSD.
10$^4$ randomly selected sets of PSDs were used to probe the
$\xsqdist$ distribution.

Figure~7 shows the best fitting singly-broken PSD model shape 
and simulated data for each
target, plotted in $f~\times~P_f$ space.
Figure~8 shows contour plots for the best-fitting singly-broken model
fixed at the best-fitting value of $\gamma$.
The presence of a constant Poisson power level dominates
the steepest PSD slopes and leads to degeneracy in that
most values of $\beta$ above $\sim$3 can
lead to the same rejection probability (e.g., upper limits on $\beta$ for 
NGC~4151 cannot be constrained); a relatively
more minor effect is that excessive red-noise leak from very steep
PSD slopes ($\beta~\gtrsim~2.5$) increases the errors,
with the result that excessively large errors decrease the
reliability of the values of the rejection probability.
For each target, the best-fitting values of $\beta$, $\gamma$, $f_{c}$, and $A$
and the likelihood of acceptance are summarized in Table~5.
As before, errors on a single parameter are 1$\sigma$ above the likelihood
of acceptance for the best-fit value. The reader can also refer to
the contour plots of
Figure~8 for estimates of the absolute rejection 
probabilities for a given set of parameters.
For Akn~564, no lower limit to $\gamma$ is given;
models with blue-noise ($\gamma$~$<$~0) PSD components  
are dominated by the aliased
power, resulting in degenerate values of the rejection 
probability for any $\gamma$~$<$~0.
The errors on $A$ are determined from the RMS spread of the 10$^4$ 
randomly selected sets of simulated PSDs.

For all six sources, the likelihood of acceptance 
improves when the break is added to the fit, as illustrated 
by the values of $\Delta$$\sigma$, the increase in 
likelihood of acceptance between the unbroken and 
singly-broken power law fits, listed in col.\ (7) of Table~5. 
This in itself indicates problems with the pure 
power-law model.  In four PSDs (Akn~564, NGC~3783, NGC~3516 and
NGC~4151), the unbroken power law fit is 
accepted at less than 10$\%$ confidence while the
singly-broken power law is 
accepted at greater than 10$\%$ confidence. 
(That is, the rejection probability drops from greater than to 
less than 90$\%$).
These are significant improvements, corresponding to an increase 
in likelihood of acceptance of 0.9--2.9$\sigma$ for these four 
sources. In the remaining two cases (Fairall~9 and NGC~5548) 
unbroken power law fits that were already reasonably acceptable 
(11.3\% and 37.2\% likelihood of acceptance)  so
the addition of a break only improves the likelihood of acceptance 
by 0.4--0.6$\sigma$.

\subsection{A doubly-broken power law in NGC~3783? }

Inspection of Figure~7 suggests that the PSD NGC~3783 is steep at high
frequencies, then flattens out in $ f \times P_f $ space (a slope of
$\approx -1 $ in $P_f $ space), then turns downward in 
$ f \times P_f $ space (to a slope 
of $\approx 0 $ in $P_f $ space) at the lowest frequencies probed.
This behavior is reminiscent of low-state
XRBs (e.g., Nowak \et\ 1999a) and suggests an improvement upon the
singly-broken fit in NGC~3783.

In order to test for a second break, we employ a model of the form

\[P(f)= \left\{ \begin{array}{ll}
                              A_l,          & f \le f_l \\
                              A_h(f/f_h)^{-1},  & f_l < f \le f_h \\
                              A_h(f/f_h)^{- \beta},  & f > f_h \end{array}
\right. \]
where $A_l$ is the PSD amplitude below the low-frequency break $f_l$,
where the PSD has zero slope.
The intermediate slope, between the two break frequencies, is fixed at --1
in order to fix this model with the same number of free parameters as the
singly-broken model.
$A_h$~=~A$_l$~$\times$$(f_h/f_l)^{-1}$ and equals the
PSD amplitude at the high frequency break $f_h$; the PSD slope
above $f_h$ is --$\beta$.
Break frequencies were tested in the log from --8.0 to --4.0 in increments
of 0.1, corresponding to $f~\rightarrow~1.26f$ in the linear scale.
The range of $\beta$ tested was 1.1--2.4 in increments of 0.1.
100 simulated PSDs were used to determine the average model PSD.
10$^4$ randomly selected sets of PSDs were used to probe the
$\xsqdist$ distribution.

The best fitting model, plotted in Figure~9, has a low-frequency
break at $f_l$ = 2.00$^{+3.01}_{-1.20}$ $\times$ 10$^{-7}$~Hz, a
high-frequency
break at $f_h$ = 3.98$^{+6.02}_{-1.47}$ $\times$ 10$^{-6}$~Hz, and a power
law slope of $\beta~=~2.0\pm0.3$ in $P_f$ space
above the high frequency break.
The best-fitting power amplitudes are 
$A_h = 2.8^{+0.3}_{-0.2} \times 10^3$~Hz$^{-1}$ and
$A_l = 5.6^{+0.6}_{-0.5} \times 10^4$~Hz$^{-1}$.
The likelihood of acceptance is 25.5$\%$ 
(rejection probability of 74.5$\%$), only a modest improvement
over the singly-broken power law model fit.
The evidence in favor of two breaks in the
NGC~3783 PSD is therefore only tentative at this point.

One can speculate, however, that the value of 
$\gamma~=~0.40^{+0.25}_{-0.35}$ in
the best-fitting singly-broken power law model
for NGC~3783 actually reflects an average of the intermediate 
and white noise portions of a doubly-broken PSD.
More significant, when $\gamma$ in the singly-broken model
is fixed to the intermediate power law slope of --1 
(similar to the best-fit singly-broken 
models for the other broad-line Seyfert~1 targets),
the best fit, with $\beta~=~1.95^{+0.15}_{-0.20}$
and $f_{c}~=~6.31^{+3.89}_{-3.80}$$\times$10$^{-6}$~Hz,
has a very low likelihood of acceptance, 2.2$\%$
(rejection probability of 97.8$\%$).

If Seyfert~1 PSDs do resemble XRB PSDs and have an an intermediate
power law slope of --1, then the PSD of NGC~3783 does require
a second break and resembles strongly the PSDs of low-state XRBs.

\subsection{Testing the stationarity of the light curves}

In order to combine the different time scale PSDs to synthesize a single
broadband PSD, we must assume that the specific short time scale PSD
measured in a single intensive \xmm\ or {\it Chandra}
observation is representative of the
average short time scale PSD that would have been measured if the source
was monitored continuously for the entire $\sim3$~yr period.
We have tested this assumption using the five {\it Chandra} scans of
NGC~3783, all taken with identical sampling intervals and durations.
Each was used to measure separate short time scale PSDs, which have
identical spectral coverage and suffer the same levels of red-noise leak
and aliasing.
This allows a straightforward comparison, as shown in Figure~10.
Note that all the PSDs show identical slopes to within the errors, and the
amplitude normalization varies by $\sim$40 percent.
This small range of amplitudes is fully
consistent with stationary behavior, providing
support for a key assumption of the PSD synthesis technique,
and consistent with the linear RMS-flux relation in the
light curves of XRBs and Seyfert~1s discussed in Uttley \& McHardy (2001).

The assumption of non-stationarity in the PSD
applies not only to the short-term data,
but must be applicable to the long-term light curves as well.
To examine the assumption of stationarity
on long time scales, we halved all
the long-term light curves and applied the Monte Carlo
analysis to the resulting PSDs by assuming the best-fitting
model for each target.
Reasonable fits were obtained for all targets, supporting
stationarity of the PSDs on time scales of
$\gtrsim$1.5 years.

\section{Discussion}

The fitting results are summarized in Figure~11, which 
allows the reader to directly assess the significance 
of features in the PSDs.  These data are plotted in 
``model space" for ease of interpretation, with the 
data shifted relative to the model as required to 
account for aliasing and other distorting effects. This 
is the opposite of the earlier figures, done in ``data space" 
in which the model is shifted to account for distortion effects.

The plot shows clearly the inadequacy of describing these 
PSDs as traditional red-noise pure power-laws. All of the 
PSDs appear to show a similar departure from power-law 
behavior, a flattening to lower temporal frequencies, 
although the break strength and frequencies differ from object 
to object.  The downturn is greatest in the PSDs of NGC~3783 (for
which a "double break" is tentatively favored) and Akn~564. 
NGC~3516 and NGC~4151 also show strong breaks.  This is 
confirmed by statistical tests, which find that addition 
of a break improves the likelihood of acceptance from less 
than 10$\%$ confidence to greater than 10$\%$ confidence 
for all four of these PSDs (see \S~4.2.). On the other hand,
the PSDs of NGC~5548 and Fairall~9 are much steeper, with a 
turnover at lower temporal frequencies, and the addition of 
a break is not statistically significant.  Only the upper 
limits to the break frequencies measured for Fairall~9 and 
NGC~5548 will be considered henceforth. We conclude that 
there is clear visual and statistical evidence of a 
significant flattening in four of these objects, and visual 
indications of a break in the remaining two.
The implications of these results are discussed below.

\subsection{Comparison with previous results}

The present analysis finds a break in the PSD of NGC~3516
at a frequency of 2~$\times$~10$^{-6}$~Hz (corresponding to
a time scale of 6 days), a factor of $\sim$5 higher
frequency than the break found by EN99 in this object.
The most likely cause for this discrepancy is the fact the EN99 forced the
power law slope below the break to zero; the present analysis
finds a much steeper low-frequency slope,
leading to the PSD break being modelled to occur at higher frequencies.

In their analysis of the Akn~564 PSD, Pounds \et\ (2001) 
also forced the low-frequency slope to zero and
also found a power law slope above the break consistent with $-1$,
consistent with the model fit in this work. However, the present work finds
a break frequency that is a factor of $\sim$2 higher;
this is most likely due to the fact that
Pounds \et\ (2001) did not account
for the large quantity of aliased power present in the PSD
and allowed the relative
normalizations of individual PSD segments to vary.

The sample of four objects in UMP02
was chosen on the basis of known strong X-ray variability,
whereas the current sample was chosen for 
2--10~keV flux (Piccinotti \et\ 1982).
The UMP02 sample could hence be biased towards rapid variability.
Two objects in the UMP02 sample, NGC~3516 and NGC~5548, overlap
with ours, and the PSD analyses yield consistent results.
The two other objects, NGC~5506 and MCG--6-30-15, are modelled
to show very high frequency breaks ($\sim$5$\times$10$^{-5}$Hz) 
when a break to a fixed low-frequency slope of --1 is assumed;
however it also should be noted that fitting these
objects' broadband PSDs with a model incorporating a fixed low-frequency
slope of 0 yields PSD breaks located an order of magnitude lower
in frequency.
NGC~5506, from the UMP02 sample, is a Seyfert~2 and will not be discussed 
further. MCG--6-30-15 will be included along with the other six Seyfert~1
galaxies in the ensuing subsections below.


In their PSD analysis, Hayashida et al.\ (1998) defined a characteristic 
variability frequency in the PSDs of AGN and Cyg X-1 as being the 
frequency at which the $ f \times P_f $ 
power crossed a certain threshold value (10$^{-3}$).  
For the PSD shapes observed 
in XRBs and Seyferts, such a frequency is dependent on 
the PSD break frequency, high-frequency slope and normalisation.  However, 
all these quantities can vary in XRBs, leading to
degenerate values of the characteristic variability frequency
for a given black hole mass.
It is also more plausible that the break frequencies, and 
not the high-frequency slopes, are more closely connected to the black hole 
mass (see $\S$~5.3).
Furthermore, the distortion effects of red-noise leak and aliasing were not
taken into account when determining those PSDs. 

Complementary to the work of Pounds \et\ (2001), 
Papadakis \et\ (2002) claim evidence for a break in PSD
slope from $\sim-1.3$  to $\sim-1.7$  above a temporal frequency 
of $\sim2~\times~10^{-3}$~Hz in Akn~564. However,
Papadakis \et\ (2002) estimate the high-frequency noise
level simply from the average flux errors. Consequently,
any small error in this noise level determination, which dominates
the highest frequencies probed, requires a subtantial change
in high-frequency PSD slope. 
Additionally, that {\it ASCA} data only were used 
leaves about 2/3 of a order of
magnitude of temporal frequency uncovered, due to Earth-occultation gaps;
more importantly, distortion effects are not considered.

\subsection{Comparison with X-ray binaries}

The Seyfert PSDs are much more poorly defined than those seen in
low-state XRBs (e.g., Nowak \et\ 1999a, Nowak \et\ 1999b).
McHardy (1988) first made the link between the two types of compact systems
based on similar slopes of the
{\it EXOSAT} power-law PSDs (see also Lawrence \&
Papadakis 1993).  The detection of Seyfert~1 PSD breaks (eg, EN99), on
time scales that appear to scale linearly with the breaks seen in XRBs,
helped confirm this picture. 
Furthermore, the energy spectra of both types of objects are similar, featuring
a non-thermal hard X-ray coronal power law which steepens
as the source brightens (e.g., Markowitz \& Edelson 2001, Wilms~et~al.~1999),
a 6.4~keV Fe~K$\alpha$
fluorescence line modelled to orignate in the regime of strong gravity
(Tanaka~et~al.~1995, Fabian~et~al.~1995, Wilms~et~al.~1999),
and a Compton reflection hump around 20--30~keV.
These spectral components, along with the rapid X-ray variability
and PSD similarities such as and the energy dependence of the PSDs
(Nandra \& Papadakis 2001)
and apparent phase lags (Papadakis, Nandra \& Kazanas 2001),
supporting common
X-ray emission and variability mechanisms in both classes
of compact objects.

The observed PSD break frequencies in Seyferts 
are about 6--7 orders of magnitude smaller than those
see in low-state XRB PSDs.
It is remarkable to note that this 
ratio is approximately the same as that of
the X-ray luminosities
and putative black hole masses (10$\Msun$ for Cyg~X-1,
Herrero \et\ 1995;
10$^{7-8} \Msun$ for Seyferts, e.g., Kaspi \et\ 2000).
This scaling supports the picture of similar variability processes
operating in both types of accreting
compact systems. 
It is noted that this type of
scaling is subject to the caveat of Seyferts and low-state XRBs 
differing {\it only} in mass, physical size, 
luminosity and variability time scale.
In fact, differences between the two classes of objects, most notably
lower inner accretion disk temperatures in Seyferts than seen
in low-state XRBs, prevent the 
analogy from being solidified.

However, key questions remain in this picture.
Are the PSD shapes really identical between
the two types of systems? The fact that
XRB PSDs do not strictly adhere to a doubly-broken
power law (e.g., GX~339--4, Nowak \et\ 1999b) and that
our analysis only quantifies the overall broad PSD shape in
terms of simple broken power laws hinder this issue.
The sharpness of these models' breaks
lacks a physical basis, but the data are not adequate enough to accurately
constrain the 'bluntness' of the break.
Do the Seyfert PSD breaks seen correspond to the high
or low frequency breaks seen in XRB PSDs? 
The tentative finding of two PSD breaks in NGC~3783 
matches closely the double breaks in XRB PSDs,
and the ratio of the break frequencies in both objects
are nearly the same ($\sim20$ for NGC~3783, $\sim$20
for Cyg~X-1 by Nowak \et\ 1999a; 25 for the ratio of
the average Cyg~X-1 break frequencies, Belloni \& Hasinger 1990),
though the low-frequency break in Cyg X-1 is
somewhat more variable and so these ratios should be taken 
with a grain of salt. 
For the other Seyferts, however, this issue is not as clear.
The singly-broken model fits for the PSDs of 
NGC~3516, NGC~4151, and to a lesser extent, NGC~5548 and Fairall~9,
appear similar to the high frequency break
in XRBs in terms of incorporating a break above which the power
law slope is $\sim$--2 and below which the slope is $\sim$--1
(see UMP02 for additional discussion).
In contrast to the broad-line targets, the best-fitting model
for the soft-spectrum source
Akn~564 seems to mimic the low-frequency breaks seen in low-state
XRBs, that from the intermediate slope of --1 above the 
break to zero below it (consistent in shape to the low-frequency
break in the doubly-broken model fit to the NGC~3783 PSD). Alternatively, 
we caution the reader that soft-spectrum 
targets like Akn~564 may in fact
exhibit fundamentally different variability characteristics 
in the 2--10 keV band 
from broad-line Seyfert~1 galaxies (see e.g., Edelson \et\ 2002), and 
the entire underlying
broadband PSD shape may in fact be significantly different in shape from
low-state XRBs. The limited 
dynamic range of the present
PSD for this target hinders this question.
The low-frequency PSD flattening to zero slope in Akn~564 presents
difficulty in solidifying the inviting analogy between 
soft-spectrum Seyfert~1 galaxies such as Akn~564 
and high-state XRBs (which show $1/f$ noise
down to frequencies of a few $\times$~$10^{-3}$~Hz,
e.g., Churazov, Gilfanov \& Revnivstev 2001), 
as both types of systems are suspected
of accreting at relatively high fractions of the Eddington 
limit (e.g., Pounds, Done \& Osborne 1995)
and both show a steep energy spectrum
($\Gamma \sim 2.4$, e.g., Vaughan \et\ 1999; Leighly 1999).
Finally, what physical process is responsible for the variability?
There are few specific models available, though some are discussed 
briefly in $\S$~5.4.

\subsection{Correlations between break time scale, luminosity and
mass estimates}

For the seven Seyfert~1s under consideration, Table~6
lists bolometric luminosity
$L_{\rm bol}$, black hole mass estimate $M_{\rm BH}$,
and PSD break time scale $ T_{\rm (days)}$
from the singly-broken power law model fits
(for NGC~3783, the high-frequency break from 
the doubly-broken model fit is used).
The time scale for MCG--6-30-15 is
from the 'high-frequency break' model used by 
UMP02. 
Bolometric luminosity is calculated using the mean bolometric correction of 
Padovani \& Rafanelli (1988), $L_{\rm bol}=27L_{\rm 2-10~keV}$.
It should be noted that for the soft-spectrum Seyfert~1 Akn~564,
the 2--10 keV X-ray to bolometric luminosity conversion of 
Padovani \& Rafanelli (1988) 
should be treated as a conservative lower limit due to the large EUV excesses
present in soft-spectrum Seyfert~1s (see, e.g., Turner \et\ 2002).
The bolometric luminosity for MCG--6-30-15
is taken from Reynolds \et\ (1997).
All reverberation-mapped masses are taken from Kaspi \et\ (2000),
except
for NGC~3516, which is from Wanders \& Horne (1994).
Akn~564 does not have a highly reliable 
reverberation-mapped mass estimate; the  
upper virial mass
estimate is $\lesssim 8 \times 10^6 \Msun$ by
Collier \et\ (2001),
based on the lag between UV continuum flux and 
Ly$\alpha$~$\lambda$1316 variations.
Reverberation mapping has not yet been done for MCG--6-30-15, 
but its black hole mass estimate may be derived as follows:
Reynolds (2000) noted that the bulge luminosity of the host
yields an approximate bulge mass of $3 \times 10^9 \Msun$.
The FWHM H$\beta$ is 1700 km/sec (Pineda \et\ 1980), which, along with
its steep energy index nearly makes it consistent with the properties
of most 'officially'-classified NLSy1s such as Akn~564.
Using the improved black hole mass-- bulge relations from 
Wandel (2002; see their Figures~3 \& 4), for which NLSy1 
black hole masses lie about an order of magnitude below 
those for broad-line AGN and normal galaxies,
the black hole mass for MCG--6-30-15 is likely to be of order
$\sim 1 \times 10^6 \Msun$, the value adopted 
henceforth in this paper.

Figure~12 shows $L_{\rm bol}$ and black hole mass plotted versus time scale.
For luminosity versus time scale (top panel), the
Pearson correlation coefficient $r$ is 0.49, with 
a probability $P_r$ of obtaining that value of $r$ by chance
of 0.27 for seven points. The bottom panel of Figure~12 shows that 
there is an apparent correlation 
between mass and time scale, with $r=0.79$, and $P_r=3.3\times10^{-2}$ for seven points.
It should be noted that for NGC~3783, the low frequency PSD break from 
the doubly-broken model fit
is not consistent with the 
above correlation. When the singly-broken model time 
scale is used for NGC~3783 (as opposed to the 
high-frequency break from the doubly-broken fit),
$r$ decreases to 0.48 ($P_r = 0.27$ for seven points) in the 
luminosity--time scale plot, and
$r$ decreases to 0.78 ($P_r = 3.9 \times 10^{-2}$ for seven points)
in the mass--time scale plot. 

A more significant difference
arises when plotting $L_{\rm 2-10~keV}$ against time scale,
using the X-ray luminosity for MCG--6-30-15 from UMP02, 
where $r$ drops to 0.27 ($P_r~=~0.56$ for seven points).
Finally, because the PSD break for MCG--6-30-15 was obtained from a fit
by UMP02
where the low frequency slope was fixed to --1 and fixing the low frequency
to 0 results in a break time scale of 2.26 days, we note that using this
time scale instead results in 
$r$~=~0.52 ($P_r~=~0.23$ for seven points) for the 
luminosity--time scale plot and
$r$~=~0.62 ($P_r~=~0.14$ for seven points) for the 
mass--time scale plot.

The apparent correlation between mass and time scale seen in Figure~12, 
stronger
than the luminosity--time scale relation, supports the validity 
of the reverberation mapping method of black hole mass estimation;
furthermore, one could speculate that mass may be more relevant than luminosity
in governing a given Seyfert's X-ray timing properties.
The mass--time scale correlation is consistent with the linear relation
$ T_{\rm (days)} = M_{\rm BH}/10^{6.5} \Msun$,
but this is certainly not conclusive with only seven objects spanning a narrow
range of luminosity and black hole mass.
The mass--time scale correlation and the similar shapes
of the PSD breaks are consistent with a picture
in which the PSDs of all the Seyfert~1s considered here
have a similar, universal broadband PSD shape (that of
low-state XRBs) in which the PSD breaks move towards lower temporal frequency 
as black hole mass increases. This is consistent with the expectation
that compact accreting black hole systems with a relative
larger black hole mass and  
larger Schwartzschild radius, $R_{\rm Sch}$, have a
a larger X-ray emission region, requiring a comparatively 
longer duration for the source to
achieve a given amplitude of flux variability and 
longer characterstic variability time scales.
This paradigm is manifested in findings of an anticorrelation
between X-ray luminosity and excess variance, 
as measured over a fixed duration (Nandra \et\ 1997,
Turner \et\ 1999, Markowitz \& Edelson 2001).

It is also interesting to note that if one extrapolates the
linear mass--time scale relation down to Cyg~X-1's mass of 10~$\Msun$,
the inverse of the resulting time scale is 3.7~Hz,
remarkably close to Cyg~X-1's mean high-frequency break of
3.3~Hz, again supporting the notion of a common variability mechanism
in both types of objects. This is also suggestive that there is only a small
departure from linearity of the mass--time scale relation
over 6--7 decades.
Furthermore, this suggests that the PSD breaks detected are
high-frequency breaks, supported by the similar changes in slope
from $\sim$--2 to $\sim$--1 (with Akn~564 being an 
exception). 

Finally, we note that the claim of a high-frequency break
at $\sim2~\times~10^{-3}$~Hz in the PSD of Akn~564
by Papadakis \et\ (2002) does not fall anywhere near
this mass--time scale relation; Papadakis \et\ (2002)
note this fact, and consider the possibility that 
the break detected may not be indicative of black hole mass.


\subsection{Correlations between PSD amplitude, luminosity, and mass
estimates}

Figure~13 shows the bolometric luminosity and black hole mass
plotted against $A$, the PSD amplitude at the break 
(in $P_f$ space)
in the singly-broken model fits
(high-frequency break from the doubly-broken fit for NGC~3783). 
Also included is MCG-6-30-15's break amplitude from UMP02,
$A = 213 \pm 8$~Hz$^{-1}$.
Under the assumption that it is a 'low-frequency' break, that of
Akn~564 is excluded.
Though there are are only six data points, there 
appears to be strong correlation between black hole mass
and PSD break amplitude, with Pearson $r~=~0.95$ ($P_r = 3.1 \times 10^{-3}$).
There is a moderate correlation between $L_{\rm bol}$ and
PSD break amplitude, with $r~=~0.72$ ($P_r = 0.10$).

These correlations reflect how the PSD, still assumed to
have the same universal shape as low-state XRBs,
not only moves downward in temporal frequency 
but also upward in power at the break 
frequency as black hole mass increases.
In a sample of nine broad-line Seyferts (including all six
broad-line Seyferts discussed thus far),
Markowitz \& Edelson (2001)
saw that the dependence of long-term 
excess variance (measured over 300~d) on
luminosity was much shallower than the dependence on short-term
(1~d) excess variance.
This indicated that the long-term excess variances likely probed 
down to temporal frequencies below
targets' respective PSD breaks. As a given PSD moves towards lower
temporal frequencies as black hole mass increases, 
the power level at the break must increase so that long-term excess 
variance stays similar. The above correlation arises mainly because
the peak power in $f~\times~P_f$ space
is broadly the same for all targets ($f~\times~P_f \sim 0.01$),
similar to that of Cyg X-1, as discussd by e.g., UMP02, not only
strengthening the notion of similar variability mechanisms in Seyferts 
and XRBs, but also
implying similar numbers of varying X-ray emitting regions.


\subsection{Testing physical time scales}

Since the observed cutoff is expected to relate to the 
physical process that generates
the variability, one can explore the relevance of various 
physical processes to the origin of the X-ray variability
by comparing physical time scale predictions to the measured PSD time scales. 
EN99 discuss the relevance of light-travel time effects
and the orbital time scale for a standard $\alpha$-disk 
(Shakura \& Sunyaev 1973), comparing them to the PSD break observed in 
NGC~3516. EN99 note that for NGC~3516, and we note for our Seyferts here,
that these time scales are both
generally too short to be directly associated with the days-to-weeks
PSD time scales.
Additionally, the radial drift/viscous time scale is usually
a few orders of magnitude longer than observed PSD time scales.
However, the thermal and acoustic variation
time scales for a thin disk
(days to weeks given the putative black hole masses
considered;
Maraschi, Molendi \& Stella 1992; Treves, Maraschi \&
Abramowicz 1988) are closest to the measured 
characterstic variability time scales. 
We hence speculate that the mechanism of X-ray variability may be tied to
physical processes which incorporate thermal
or acoustic thin-disk variations.
If this is indeed the case, then the characteristic
time scales and black hole mass
estimates can be used to constrain the X-ray emitting
location.
Assuming reasonable values for the scale height and viscosity,
and assuming that the thin disk approximation holds
over the entire radius range,
all of the targets' acoustic and thermal time scales are consistent with
the variable X-ray emission originating within $\sim$30$R_{\rm Sch}$.

However, more specific models can be considered.
For instance, in the pulse avalanche model of Poutanen \& Fabian (1999),
the variable X-ray emission is produced in
an inhomogeneous, stochastic system of 
magnetic flares inflating and detaching
above the inner accretion disk, producing X-ray flares
via inverse-Compton scattering of softer photons, but with events occurring
in correlated avalanches. The resulting light curves can yield
doubly broken PSDs similar to those of low-state XRBs for
appropriately tuned parameters, with a
break between $f^{-\beta}$ noise and $f^{-1}$ noise
being associated with the longest time scale (days to weeks for the
broad-line Seyfert~1s considered here) of an individual flare.
A low-frequency PSD break from $f^{-1}$ to white noise is associated with
the duration of the avalanches. However, predicted
flare durations are generally too short to be directly associated with
PSD breaks. 

Along another vein, 
the X-ray variability in the model of
Lyubarskii (1997), is caused by variations in 
the accretion flow (e.g., due to turbulence)
propagating from large radii inwards
to the location of the X-ray corona.
Such fluctuations in viscosity are dependent on the local viscous time scale,
which decreases with decreasing radius.
For standard thin disks, the viscous
timescale is too long to be
associated with the measured PSD breaks.  However, if the accretion flow
is geometrically thick, such as an advection-dominated accretion flow
(ADAF; Narayan \& Yi 1994) or 
a radiation-supported thick disk (e.g. see Abramowicz 1988,
Treves, Maraschi \& Abramowicz 1988)
the viscous time-scale tends toward the 
thermal time-scale.  Therefore the PSD breaks seen might 
plausibly correspond to a viscous time-scale in an ADAF or 
radiation-supported thick disk.


\section{Conclusions}

We have systematically constructed high-quality
broadband PSDs for a sample 
of six Seyfert~1 galaxies.
These high-quality PSDs cover exceptional dynamic ranges, 
continuously spanning up to or beyond 3--4 orders of magnitude in
temporal frequency. 
We use the Monte Carlo technique of UMP02
to determine adequate errors
on each binned PSD point and to account for distortion effects, and we 
characterize the underlying PSD shape to look for
low-frequency flattening below a characteristic break frequency.
Four targets (Akn~564, NGC~3783, NGC~3516 and NGC~4151)
show significant evidence for low-frequency flattening, with
characteristic variability
time scales in the range of several to tens of days.
For the two most massive and luminous targets in 
the sample (Fairall~9 and NGC~5548),  
we expect that continued long-term monitoring
and extending the PSD to probe lower frequencies will
confirm their break frequencies, which are the lowest
in the sample.
The low-frequency flattening seen in these objects is
remarkably similar to what is seen in low-state
X-ray binary PSDs, strengthening the argument that similar emission
processes occur in both types of compact accreting systems, spanning 
a factor of $\sim$10$^{6-7}$ in luminosity and putative 
black hole mass. 

All of the PSDs studied are consistent in shape with 
at least portions of low-state XRB PSDs.
The finding of two breaks in the PSD of NGC~3783 is only tentative,
but it still heralds a new era in Seyfert PSD analysis: those
PSDs derived from {\it EXOSAT} data
generally lacked the temporal frequency range to reveal even a single
break, and the access to long time scales afforded by \xte\
allowed detection of a single break (EN99, Pounds \et\ 2001, UMP02, 
this work). However, as double power law
breaks are routinely detected in
XRB PSDs, only Seyfert PSDs which also show two breaks 
will exhibit the most direct and convincing evidence for a link between
the two classes of systems. More long-term monitoring is
therefore crucial for locating low-frequency PSD breaks.

The PSD break frequencies detected are model-dependent, but 
when compared to
reverberation-mapped black hole masses, they
are consistent with a linear 
mass--time scale relation that extends not only throughout
the Seyfert range, but down to the mass and PSD time scale of Cyg~X-1,
providing even further support for similar variability processes
in Seyferts and XRBs. 
The correlation, however, is hampered by small-number statistics,
and more accurate black hole mass estimates for Akn~564 and
MCG--6-30-15 are needed to clarify the situation;
this is especially important for MCG--6-30-15
since its mass estimate is very crude, yet it is used to anchor the low-mass
end of the black hole mass--time scale relation shown in Figure~12.

It is encouraging to note that this agreement
between X-ray variability time scale and 
reverberation mapping arises
despite numerous caveats and assumptions
in both analysis 
methods (e.g., the assumption of broad-line emission clouds being
on Keplerian orbits, Peterson \& Wandel 1999), each of which is expected
to probe different regions of the central engine.


Finally, we draw the reader's attention back to Figure~11, as we 
believe such diagrams will play a large role in future Seyfert 1 
PSD analysis.  It is a plot of $ f \times P_f $ as a function of 
$f$ in model space, while most previous PSD plots were of $P_f$ as a 
function of $f$ in data space.  In this regard, it mimics the 
``unfolded" (model-space) $ \nu \times F_\nu $ plots of early X-ray 
energy spectral analyses.  That was done to give the clearest possible 
visual picture of the instrinsic energy spectrum itself, undistorted 
by the vagaries of the response matrix of a particular detector.  
Likewise, this plot gives the clearest possible picture of the 
underlying PSD, after removing the distortions unique to the 
particular sampling pattern employed.  Further, just as a peak in the 
$ \nu \times F_\nu $ plots indicated a region of maximum luminosity 
per decade of frequency, so a well-defined peak in the $ f \times 
P_f $ PSD plot occurs in the temporal frequency band at which most of 
the variability power is produced.  These data indicate that the 
``characteristic variability time scales" of these Seyfert~1s are
typically of order a few days.
 
Establishing that breaks are a systematic feature of their PSDs 
represents a milestone in the description of Seyfert~1 X-ray 
variability.  Here too there is analogy with the early energy spectral 
studies: the low-resolution energy spectra from early missions 
(e.g., Ariel V) were well-modeled as ``canonical" $ \Gamma \sim 
1.7 $ power-laws, but a critical breakthrough came when Wilkes \& 
Elvis (1987) unambiguously established the existence of a significant 
 departure, the ``soft excess," from the {\it Einstein} data.  As the 
spectral resolution and bands improved over the following 15 years, 
many more spectral features have been found (e.g., narrow
absorption lines that make up the ``warm absorber," the iron K$_\alpha$
fluorescence line, the Compton reflection component).  Now this work 
has clearly established the pervasiveness of breaks in Seyfert~1 
PSDs that were modeled as simple power-laws in previous, lower 
dynamic range PSDs. We can not be certain that this is strictly a 
broken power-law; many other parameterizations are of course possible. 
We can be fairly confident, however, that the current description of 
Seyfert~1 PSDs will be refined and eventually overturned, just as 
the ``canonical" energy spectral power-law was 15 years ago, as 
we further probe long time scales with \rxte\ and short time scales 
with \xmm, and the flight of {\it Lobster-ISS} \footnote{See http://www.src.le.ac.uk/lobster} starting in 2009 
increases the number of Seyfert 1 PSDs more than tenfold.

\acknowledgments
The authors thank the anonymous referee for a very detailed reading of
the manuscript.
The authors also acknowledge the dedication of the entire \xte\ mission team,
especially Evan Smith, for scheduling the long-term observations so evenly
all these years. A.M.\ thanks Rick Paik-Schoenberg
of the U.C.L.A.\ Statistics Department for useful discussions on spectral
windowing. This work has made use of data obtained through the High Energy
Astrophysics Science Archive Research Center Online Service, provided by
the NASA Goddard Space Flight Center, the TARTARUS database, which is
supported by Jane Turner and Kirpal Nandra under NASA grants NAG~5-7385
and NAG~5-7067, and the NASA$/$IPAC Extragalactic Database which is
operated by the Jet Propulsion Laboratory, California Institute of
Technology, under contract with the National Aeronautics and Space
Administration. A.M.\ \& R.E.\ acknowledge financial 
support from NASA grant NAG~5-9023. R.E.G.\ acknowledges 
financial support from NASA grant NAG~5-9902.

\clearpage

\begin{deluxetable}{lcccc}
\tablewidth{4.5in}
\tablenum{1}
\tablecaption{Source Parameters for the Seyfert Galaxies \label{tab1}}
\tablehead{
\colhead{Source} & \colhead{$L_{2-10 keV}$} & \colhead{ } & \colhead{FWHM H$\beta$} & \colhead{}\\
\colhead{Name} & \colhead{(log10(erg s$^{-1}$))} & \colhead{z} & \colhead{(km sec$^{-1}$)} & {$\Gamma_{2-10}$} }
\startdata
Fairall~9 & 43.97 & 0.047 & 5900 & 2.18 \\
NGC~5548  & 43.50 & 0.017 & 5610 & 1.79 \\
Akn~564   & 43.49 & 0.024 & 720  & 2.68 \\
NGC~3783  & 43.22 & 0.010 & 2980 & 1.77 \\
NGC~3516  & 42.86 & 0.009 & 6800 & 1.60 \\ 
NGC~4151  & 42.62 & 0.003 & 5000 & 1.65 \\
\enddata
\tablecomments{Targets are ranked by unabsorbed 2--10~keV luminosity, 
listed in Col.\ (2). Luminosities
were calculated with the online {\sc W3PIMMS} tool and 
using the long-term mean \xte\ count rates,
X-ray photon indices, assuming a cold absorption
column equal to the Galactic column (for NGC~4151, a column
density 230$\times$ the Galactic column was used;
see Schurch \& Warwick 2002),
and calculated assuming $H_{0}=75$~km~s$^{-1}$~Mpc$^{-1}$
and $q_{0} =0.5 $.
Redshifts (col.~[3]) were obtained from the NED database. 
H$\beta$ values (col.~[4]) were taken from Turner~et~al.~(1999);
see references therein. The photon indicies in col.\ (5) were obtained from the
Tartarus database, derived from a simple power law fit covering
the 2--10~keV bandpass, except for Akn~564, from a spectral fit
by Vaughan \et\ (1999),
NGC~3783, from a spectral fit to {\it Chandra} data by Kaspi~et~al.\ (2001), and
NGC~4151, from a spectral fit to \xmm\ data by Schurch \& Warwick (2002).
}
\end{deluxetable}


\begin{deluxetable}{lcccccccc}
\tabletypesize{\footnotesize}
\tablewidth{6.5in}
\tablenum{2}
\tablecaption{Sampling Parameters \label{tab2}}
\tablehead{
\colhead{Source} & \colhead{Time} & \colhead{} & \colhead{MJD Date} &  \colhead{} 
& \colhead{} & \colhead{Fraction} & \colhead{} & \colhead{$F_{\rm var}$}     \\        
\colhead{Name}  & \colhead{Scale} & \colhead{Instrument} & \colhead{Range} & \colhead{$\Dtsamp$} 
& \colhead{Npts} & \colhead{missing ($\%$)} & \colhead{$\mu$} & \colhead{($\%$) }  } 
\startdata
Fairall~9 & Long   & \xte\  & 51180.59--52311.18    & 4.27 d    & 266  & 7.9   & 1.54  & 39.2 $\pm$ 1.8 \\
Fairall~9 & Medium & \xte\  & 52144.88--52178.97    & 3.2 hr    & 258  & 13.2  & 2.09  & 14.7 $\pm$ 0.7 \\ \hline
NGC~5548  & Long   & \xte\  & 50208.07--52417.02    & 14 d    & 158  & 0.0   & 4.82  & 34.5 $\pm$ 0.9 \\
NGC~5548  & Medium & \xte\  & 52091.70--52125.40    & 3.2 hr    & 255  & 8.2   & 4.51  & 26.3 $\pm$ 1.2 \\
NGC~5548 & Short & {\it XMM}  &  52097.68--52098.63 & 900~s  & 93 & 0.0   & 2.99  & 5.3 $\pm$ 0.4  \\  \hline
Akn~564   & Long   & \xte\  & 51179.58--52310.19    & 4.27 d     & 266 & 7.9   & 1.73  & 29.2 $\pm$ 1.3 \\
Akn~564   & Medium & \xte\  & 51694.86--51726.51    & 3.2 hr     & 239 & 13.0  & 2.00  & 34.9 $\pm$ 1.7 \\ \hline
NGC~3783 & Long   & \xte\  & 51180.55--52311.25     & 4.27 d     & 266 & 9.4   & 7.09  & 22.3 $\pm$ 1.0 \\
NGC~3783 & Medium & \xte\  & 51960.17--51980.05     & 3.2 hr     & 151 & 7.9   & 5.65  & 12.8 $\pm$ 0.8 \\
NGC~3783 & Short  & \chandra\ & 51964.80--51966.74 & 2000~s   &  84 &  0.0  & 0.39  & 12.3 $\pm$ 0.9 \\
NGC~3783 & Short  & \chandra\ & 51967.40--51969.38 & 2000~s   &  84 &  0.0  & 0.39  & 11.9 $\pm$ 0.9 \\
NGC~3783 & Short  & \chandra\ & 51978.04--51979.94 & 2000~s   &  84 &  0.0  & 0.41  & 11.5 $\pm$ 0.9 \\
NGC~3783 & Short  & \chandra\ & 51999.17--52001.11 & 2000~s   &  84 &  0.0  & 0.58  & 15.1 $\pm$ 1.2 \\
NGC~3783 & Short  & \chandra\ & 52086.43--52088.38 & 2000~s   &  84 &  0.0  & 0.49  & 13.5 $\pm$ 1.0 \\ \hline
NGC~3516 & Long   & \xte\  & 50523.03--51593.40     &  4.27 d    &  252 & 15.1 & 3.77  & 37.8 $\pm$ 1.8 \\
NGC~3516 & Medium & \xte\  & 50523.03--50659.09     & 12.8 hr    &  256 & 22.7 & 4.21  & 28.0 $\pm$ 1.4 \\
NGC~3516 & Short & \xte\  & 50590.01--50594.22  & 1200~s   & 304  & 15.1 & 3.85  & 7.6 $\pm$ 0.3 \\
NGC~3516 & Short & \xte\  & 50916.34--50919.67  & 1200~s   & 241  & 29.1 & 5.09  & 10.4 $\pm$ 0.6  \\  \hline
NGC~4151 & Long & \xte\ & 51179.56--51964.65       & 4.27 d     &  185 & 6.0  & 14.86 & 38.1 $\pm$ 2.0   \\
NGC~4151 & Medium & \xte\ & 51870.60--51904.79     & 3.2 hr     & 259  & 11.2 & 7.43  & 18.4 $\pm$ 0.9  \\
NGC~4151 & Short & {\it XMM}  & 51900.48--51901.14 & 1100~s & 53   & 0.0  & 3.04  & 7.8 $\pm$ 0.8    \\
\enddata
\tablecomments{Targets are ranked by 2--10~keV luminosity. Col.\ (5), $\Dtsamp$, is the sampling interval.
Col.\ (8), $\mu$, is the mean count rate.
\xte\ count rates are per 1 PCU.}
\end{deluxetable}


\begin{deluxetable}{lccccc}
\tablewidth{5.9in}
\tablenum{3}
\tablecaption{PSD Measurement Parameters \label{tab3}}
\tablehead{
\colhead{Source} & \colhead{Time} & \colhead{$\fmin$} & \colhead{$\fNyq$} & \colhead{Temp.\ Freq.\ Range}
& \colhead{$P_{Psn}$}  \\
\colhead{Name} & \colhead{Scale} & \colhead{(Hz)} & \colhead{(Hz)} & \colhead{Spanned (decades)} & \colhead{(Hz$^{-1}$)} }
\startdata
Fairall~9 & Long   &  $1.0 \times 10^{-8}$  & $1.4 \times 10^{-6}$  &     3.6     & 1360\\
Fairall~9 & Medium &  $3.4 \times 10^{-7}$  & $4.4 \times 10^{-5}$  &                    & 30.9 \\ \hline
NGC~5548  & Long   &  $5.2 \times 10^{-9}$  & $1.4 \times 10^{-6}$  &     5.0     &  577\\
NGC~5548  & Medium &  $3.4 \times 10^{-7}$  & $4.4 \times 10^{-5}$  &                    &  9.4\\ 
NGC~5548 & Short &    $1.2 \times 10^{-5}$  & $5.6 \times 10^{-4}$  &                    &  0.69\\ \hline  
Akn~564   & Long   &  $1.0 \times 10^{-8}$  & $1.4 \times 10^{-6}$  &     3.6     &  1050\\
Akn~564   & Medium &  $3.7 \times 10^{-7}$  & $4.4 \times 10^{-5}$  &                    &  33.8\\  \hline 
NGC~3783 & Long   &   $1.0 \times 10^{-8}$  & $1.4 \times 10^{-6}$  &     4.4     &  1210\\
NGC~3783 & Medium &   $5.8 \times 10^{-7}$  & $4.4 \times 10^{-5}$  &                    &  6.87\\ 
NGC~3783 & Short  &   $5.9 \times 10^{-6}$  & $2.5 \times 10^{-4}$  &                    &  5.12\\ 
NGC~3783 & Short  &   $5.9 \times 10^{-6}$  & $2.5 \times 10^{-4}$  &                    &  5.14\\ 
NGC~3783 & Short  &   $5.9 \times 10^{-6}$  & $2.5 \times 10^{-4}$  &                    &  4.94\\ 
NGC~3783 & Short  &   $5.9 \times 10^{-6}$  & $2.5 \times 10^{-4}$  &                    &  3.45\\ 
NGC~3783 & Short  &   $5.9 \times 10^{-6}$  & $2.5 \times 10^{-4}$  &                    &  4.13\\  \hline
NGC~3516 & Long   &   $1.1 \times 10^{-8}$  & $1.4 \times 10^{-6}$  &     4.6     &  177\\
NGC~3516 & Medium &   $8.5 \times 10^{-8}$  & $1.1 \times 10^{-5}$  &                   & 13.8\\
NGC~3516 & Short &    $2.8 \times 10^{-6}$  & $4.2 \times 10^{-4}$  &                   & 0.37\\
NGC~3516 & Short &    $3.5 \times 10^{-6}$  & $4.2 \times 10^{-4}$  &                   & 0.23\\ \hline
NGC~4151 & Long &     $1.1 \times 10^{-8}$  &  $1.4 \times 10^{-6}$  &       4.6        &  44.2  \\
NGC~4151 & Medium &   $3.4 \times 10^{-7}$  & $4.4 \times 10^{-5}$  &                   &  4.7 \\ 
NGC~4151 & Short  &    $1.2 \times 10^{-5}$  & $4.5 \times 10^{-4}$  &                  &  0.67\\ 
\enddata
\end{deluxetable}

\begin{deluxetable}{lccc}
\tablewidth{4.8in}
\tablenum{4}
\tablecaption{Results For Unbroken Power Law Model Fit \label{tab4}}
\tablehead{
\colhead{} & \colhead{Best-} & \colhead{Best-Fitting $A_o$} & \colhead{Likelihood} \\
\colhead{Target} & \colhead{Fitting $\beta$} & \colhead{(units of Hz$^{-1}$)} &  \colhead{of acceptance ($\%$)} }
\startdata
Fairall~9 & 1.60$^{+2.40}_{-0.20}$  & 2.7$\pm0.4 \times 10^3$           &  11.3\\
NGC~5548 &  1.65$^{+0.20}_{-0.10}$  & 3.4$^{+0.6}_{-0.5} \times 10^3$   &  37.2\\
Akn~564  &  0.75$^{+0.40}_{-0.10}$  & 4.5$\pm0.5 \times 10^3$           &  0.6\\
NGC~3783 &  1.25                    & 5.5$\pm0.5 \times 10^3$           &  $<$0.01\\
NGC~3516 &  1.80$^{+2.20}_{-0.35}$  & 9.3$^{+1.6}_{-1.9} \times 10^3$   &  6.6\\
NGC~4151 &  1.90$^{+2.10}_{-0.35}$  & 7.8$^{+3.4}_{-2.4} \times 10^3$   &  1.8\\
\enddata
\tablecomments{Results from fitting the PSDs with an unbroken power law model.
$A_o$ is the power amplitude at $f~=~10^{-6}$~Hz.
The likelihood of acceptance is defined as (1 -- $R$), where $R$ is the rejection confidence.
No error is assigned for the best-fitting $\beta$ for NGC~3783.
Upper limits on $\beta$ for Fairall~9, NGC~3516 and NGC~4151 cannot be constrained
due to a red-noise leak bias discussed in the text.}
\end{deluxetable}

\begin{deluxetable}{lcccccc}
\tablewidth{6.3in}
\tablenum{5}
\tablecaption{Results For Singly-Broken Power Law Model Fits \label{tab5}}
\tablehead{       
\colhead{}       & \colhead{}         & \colhead{}        & \colhead{}             & \colhead{}       & \colhead{Likelihood}   & \colhead{}       \\
\colhead{Target} & \colhead{$\gamma$} & \colhead{$\beta$} & \colhead{$f_{c}$ (Hz)} &  \colhead{$A$ (Hz$^{-1}$)} & \colhead{of acceptance ($\%$)}   & \colhead{$\Delta$$\sigma$}      }
\startdata
Fairall~9  &  1.10$^{+1.10}_{-0.60}$ & 2.20$^{+0.65}_{-0.20}$ & 3.98$^{+2.33}_{-2.40}$$\times$10$^{-7}$  & 5.4$^{+0.8}_{-0.7} \times 10^4$ & 23.5 & 0.6 \\
NGC~5548   &  1.15$^{+0.50}_{-0.65}$ & 2.05$^{+0.80}_{-0.40}$ & 6.31$^{+18.8}_{-5.05}$$\times$10$^{-7}$  & 2.5$^{+0.5}_{-0.4} \times 10^4$ & 87.4 & 0.4 \\
Akn~564    &  0.05$^{+0.55}_{-2.05}$ & 1.20$^{+0.25}_{-0.35}$ & 1.59$^{+4.73}_{-0.95}$$\times$10$^{-6}$  & 1.9$^{+0.1}_{-0.2} \times 10^4$  &  97.3 & 2.9 \\
NGC~3783   &  0.40$^{+0.25}_{-0.35}$ & 1.90$^{+0.15}_{-0.30}$ & 2.00$^{+3.01}_{-0.74}$$\times$10$^{-6}$  & 1.0$\pm0.1 \times 10^4$  & 12.7  & 2.0 \\
NGC~3516   &  1.10$^{+0.40}_{-0.30}$ & 2.00$^{+0.55}_{-0.20}$ & 2.00$^{+3.01}_{-1.00}$$\times$10$^{-6}$  & 7.9$^{+0.8}_{-0.7} \times 10^3$ & 81.1 & 1.9\\ 
NGC~4151   &  1.10$^{+1.35}_{-0.40}$ & 2.10$^{+1.90}_{-0.25}$ & 1.26$^{+1.90}_{-1.01}$$\times$10$^{-6}$  & 1.3$\pm0.2 \times 10^4$ & 10.8 & 0.9 \\ 
\enddata
\tablecomments{Results from fitting the PSDs with the singly-broken power law  model.
$\gamma$ is the low frequency power law slope, $\beta$ is
the high-frequency power law slope, and $A$ is the power amplitude at
the break frequency $f_{c}$. The likelihood of acceptance is again
defined as (1 -- $R$), where $R$ is the rejection confidence.
$\Delta$$\sigma$ quantifies the increase in 
likelihood of acceptance between the unbroken and singly-broken 
power law fits.}
\end{deluxetable}

\begin{deluxetable}{lccc}
\tablewidth{4.8in}
\tablenum{6}
\tablecaption{Time Scale, Bolometric Luminosity, and Black Hole Mass Estimate}
\tablehead{
\colhead{} & \colhead{PSD Break} & \colhead{$L_{\rm bol}$} & \colhead{B.H. Mass} \\
\colhead{Target} & \colhead{Time Scale (days)} & \colhead{log(erg s$^{-1}$)} & \colhead{ ($10^7$$\Msun$) } }
\startdata
Fairall~9    & $>$18.3                & 45.61 & 8.3$^{+2.5}_{-4.3}$ \\
NGC~5548     & $>$4.6                 & 45.01 & 9.4$^{+1.7}_{-1.4}$ \\
Akn~564      & 7.3$^{+11}_{-5.5}$   & 44.81 & 0.8\\
NGC~3783     & 2.9$^{+2.9}_{-1.4}$    & 44.65 & 1.10$^{+1.07}_{-0.98}$ \\
NGC~3516     & 5.8$^{+5.8}_{-3.5}$    & 44.36 & 2.0$\pm$0.3 \\
MCG--6-30-15  &  0.23$^{+0.67}_{-0.12}$       & 44.24 & 0.1 \\
NGC~4151     & 9.2$^{+37}_{-5.5}$     & 44.06 & 1.20$^{+0.83}_{-0.70}$\\
\enddata
\tablecomments{Targets are ranked by L$_{\rm Bol}$.
See text for details.}
\end{deluxetable}

\clearpage
\begin{figure}[!hb]
\epsscale{0.83}
\plotone{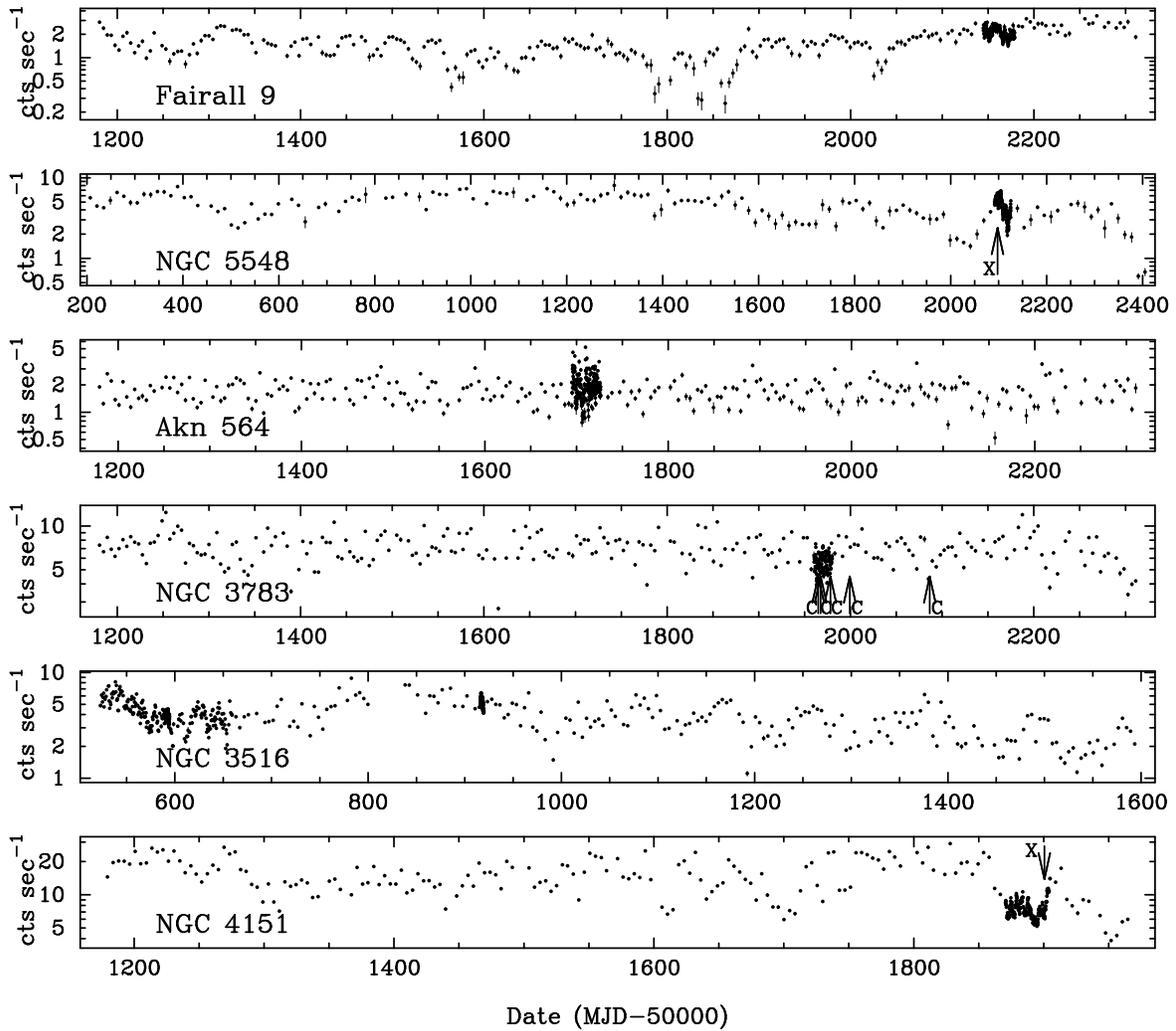}
\caption{2--10~keV {\it RXTE}
'raw' light curves
for the six targets. 'X' and 'C' denote location of
\xmm\ and \chandra\ long-looks respectively.
Count rates are all normalized to 1~PCU$^{-1}$.}
\end{figure}

\begin{figure}[hb]
\epsscale{0.83}
\plotone{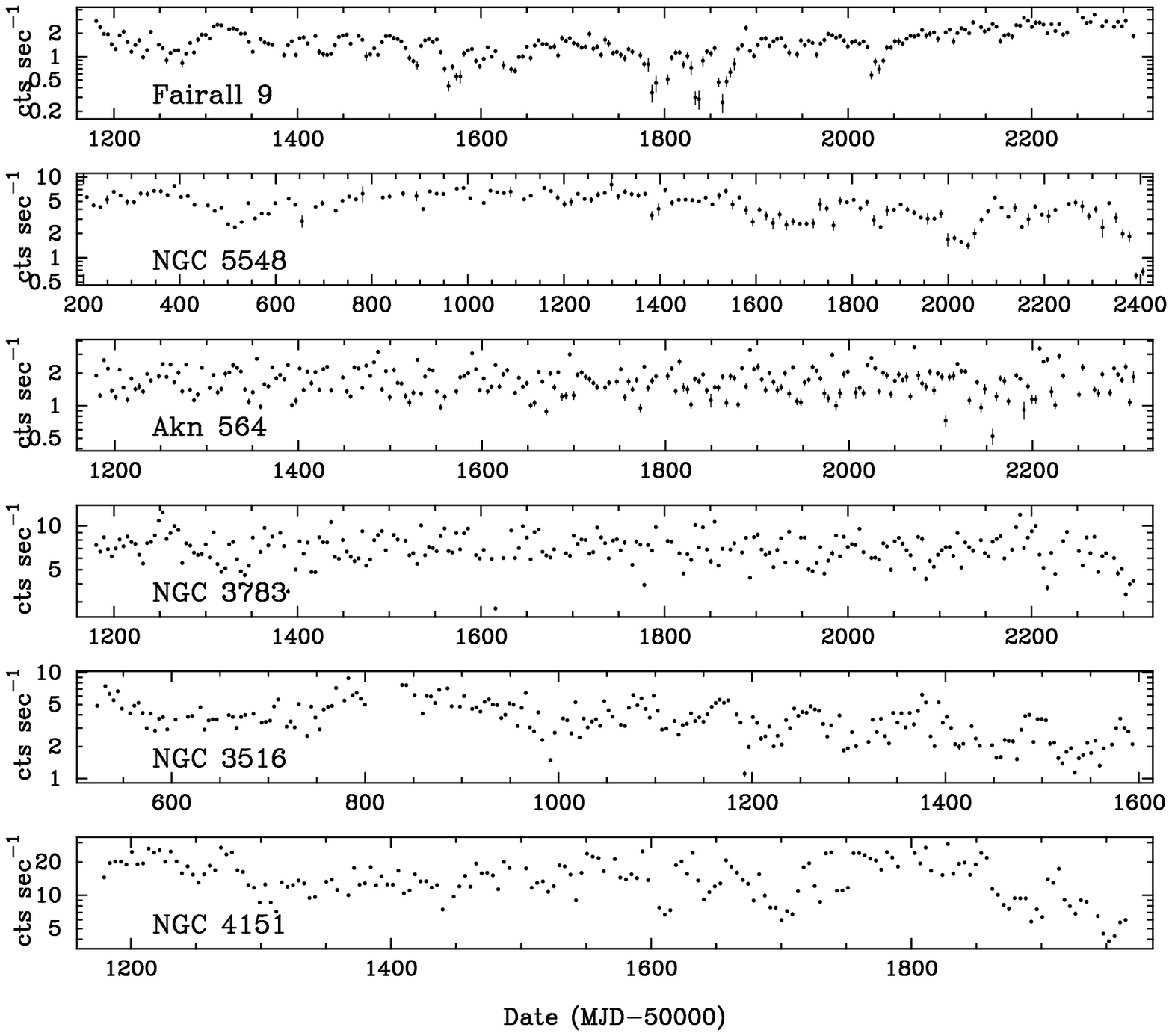}
\caption{2--10~keV {\it RXTE}
long-term monitoring light curves
for the six targets. Count rates are all normalized to 1~PCU$^{-1}$.}
\end{figure}

\begin{figure}[hb]
\epsscale{0.83}
\plotone{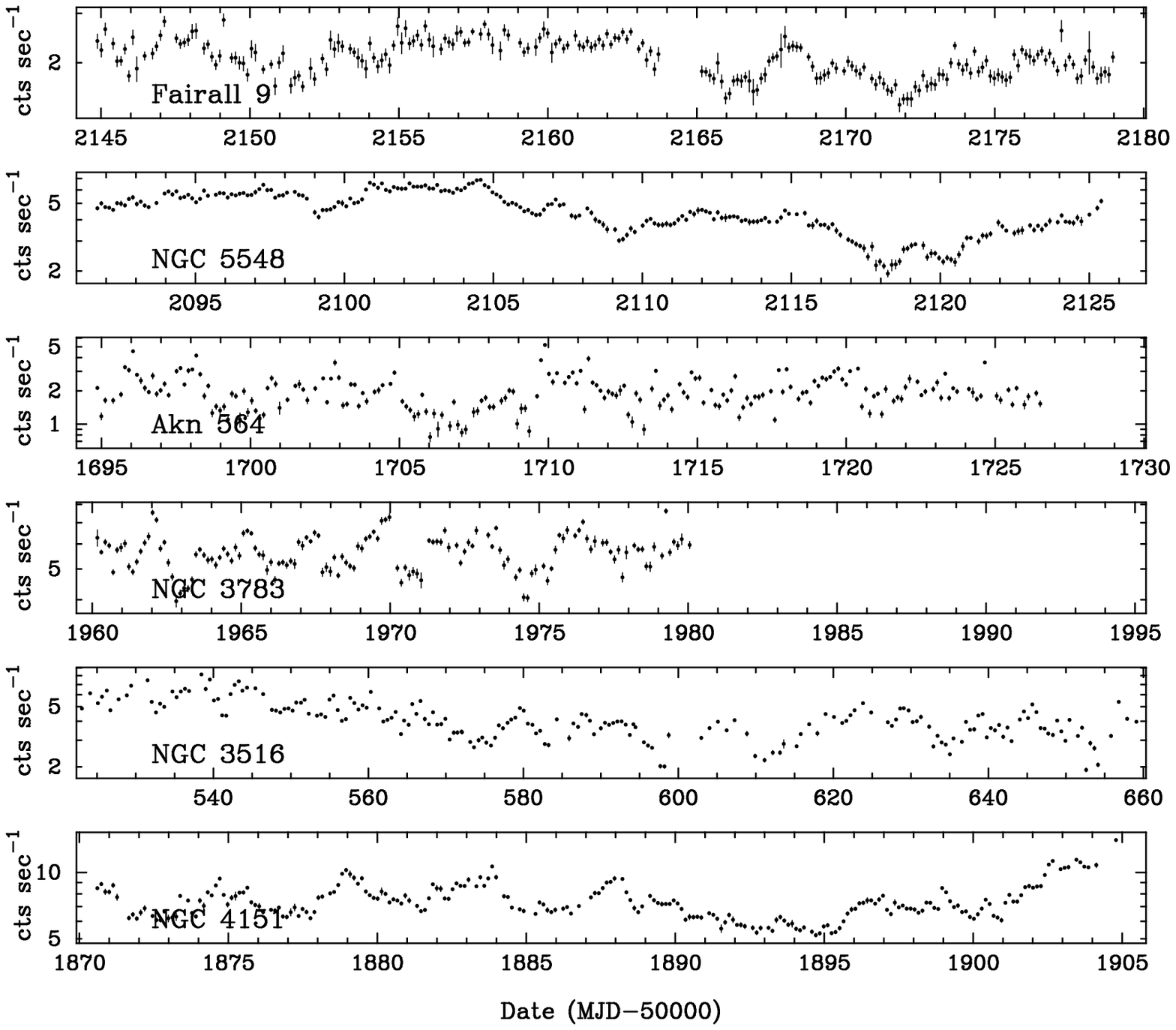}
\caption{2--10~keV {\it RXTE}
medium-term monitoring light curves
for the six targets. Count rates are all normalized to 1~PCU$^{-1}$.}
\end{figure}

\begin{figure}[hb]
\epsscale{0.77}
\plotone{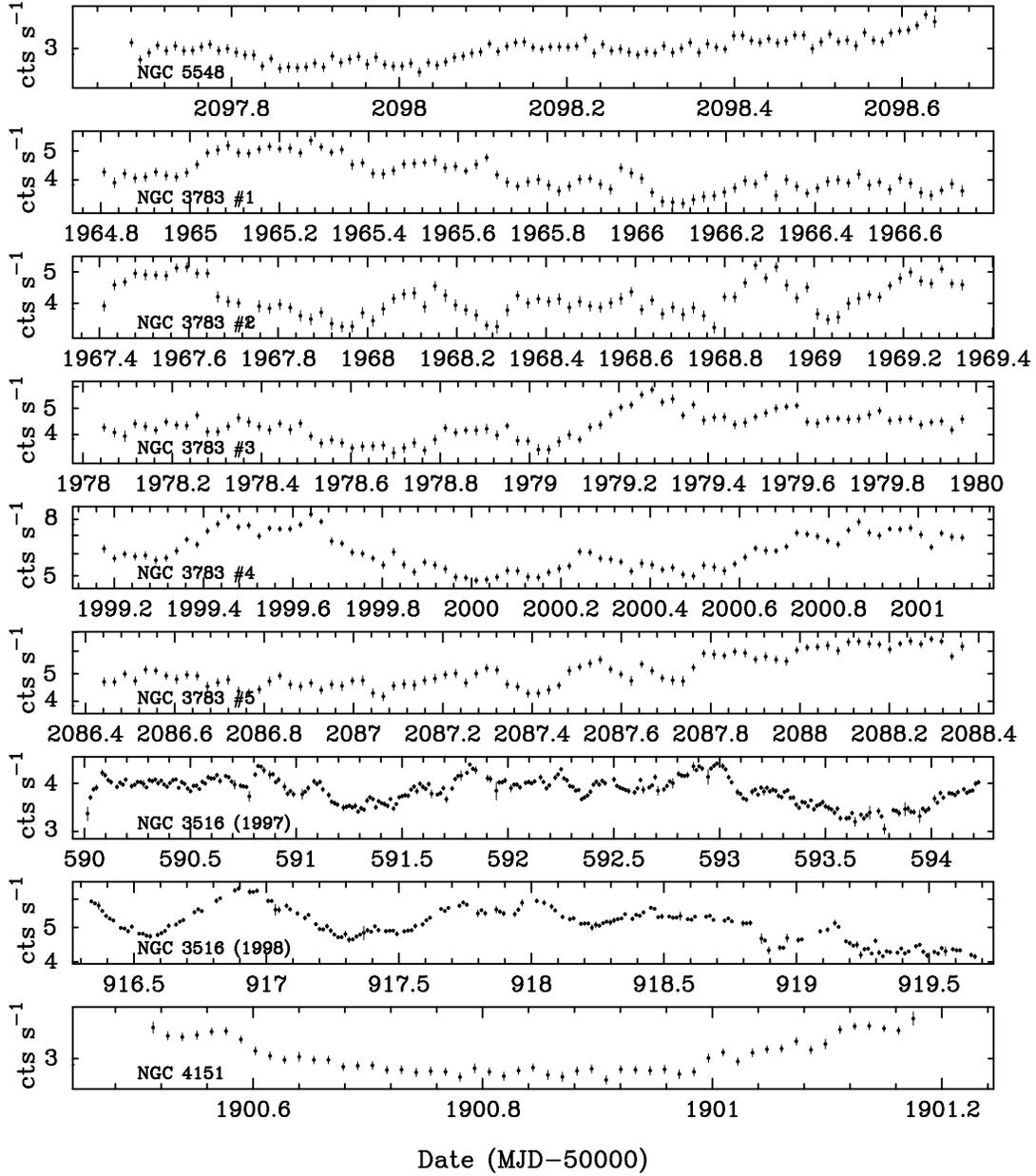}
\caption{2--10~keV continuous long-look light 
curves. Top to bottom are: NGC~5548 with {\it XMM-Newton}; the five
{\it Chandra} long-looks for NGC~3783; the 1997 ($\#$1) and 1998 ($\#$2)
{\it RXTE} long-looks for NGC~3516; and NGC~4151 with {\it XMM-Newton}.
Count rates have not been renormalized relative to each other.
PCA count rates are all normalized to 1~PCU$^{-1}$.}
\end{figure}

\begin{figure}[hb]
\epsscale{0.95}
\plotone{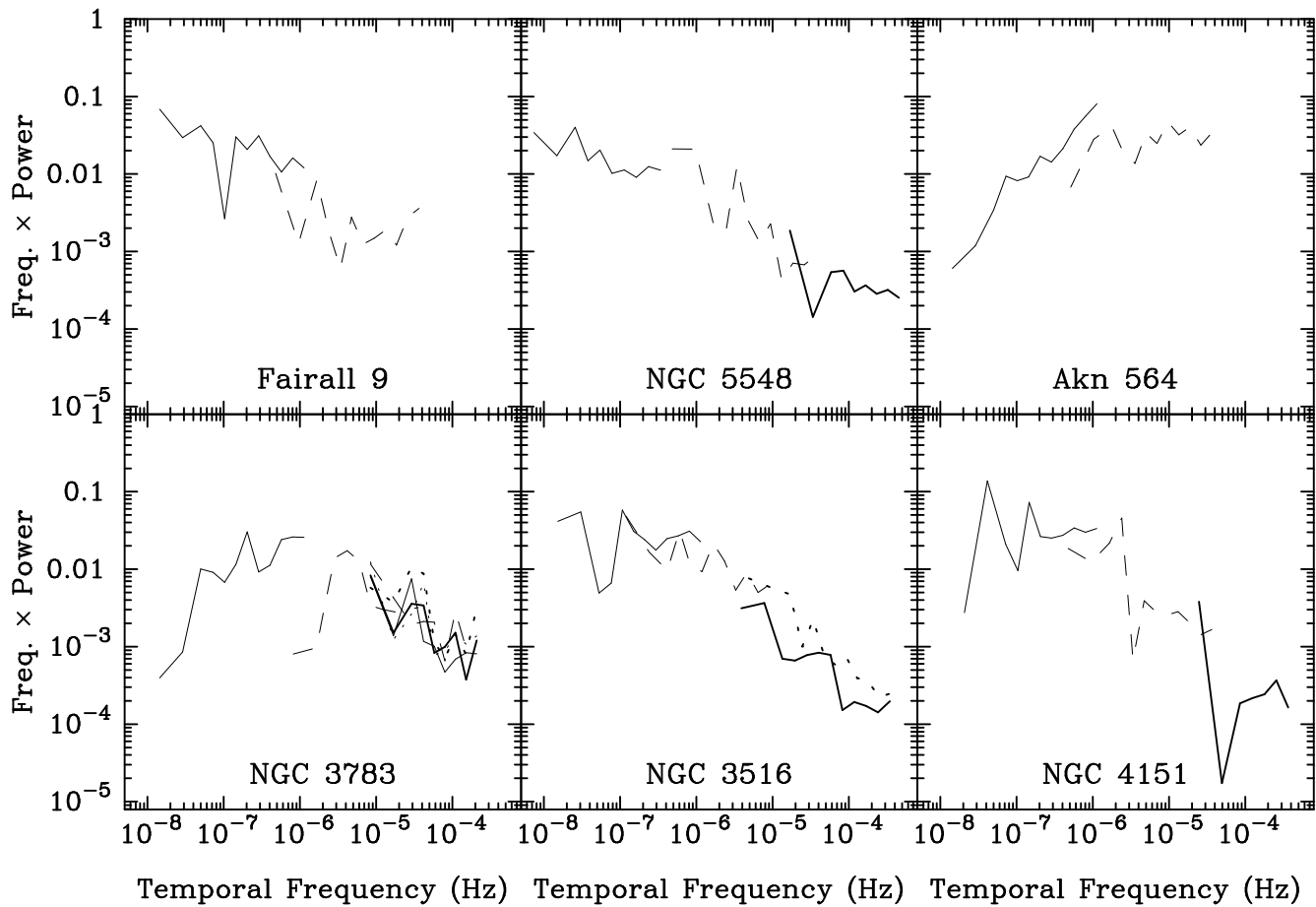}
\caption{ Raw broadband PSDs for the six targets, plotted 
in $f~\times~P_f$ space (units are Hz~$\times$~Hz$^{-1}$, or dimensionless).
Such $f~\times~P_f$ plots are common in XRB PSD analysis; e.g., Sunyaev \& Revnivtsev 2000). Targets are ranked by 2--10~keV luminosity; the power due
to Poisson noise has not been subtracted from these PSDs.
Each long-term PSD is marked by a solid line; medium-term PSD with
dashed line. The short-term NGC~5548 and NGC~4151 PSDs are
marked with a solid line. The NGC~3516 1997 and 1998 short-term
PSDs are marked with solid and dotted lines, respectively.
The NGC~3783 short-term \chandra\ PSDs, from first- to last-observed,
are denoted by bold solid, 
bold dotted, solid, dashed, and dotted lines,
respectively.}
\end{figure}

\begin{figure}[hb]
\epsscale{0.95}
\plotone{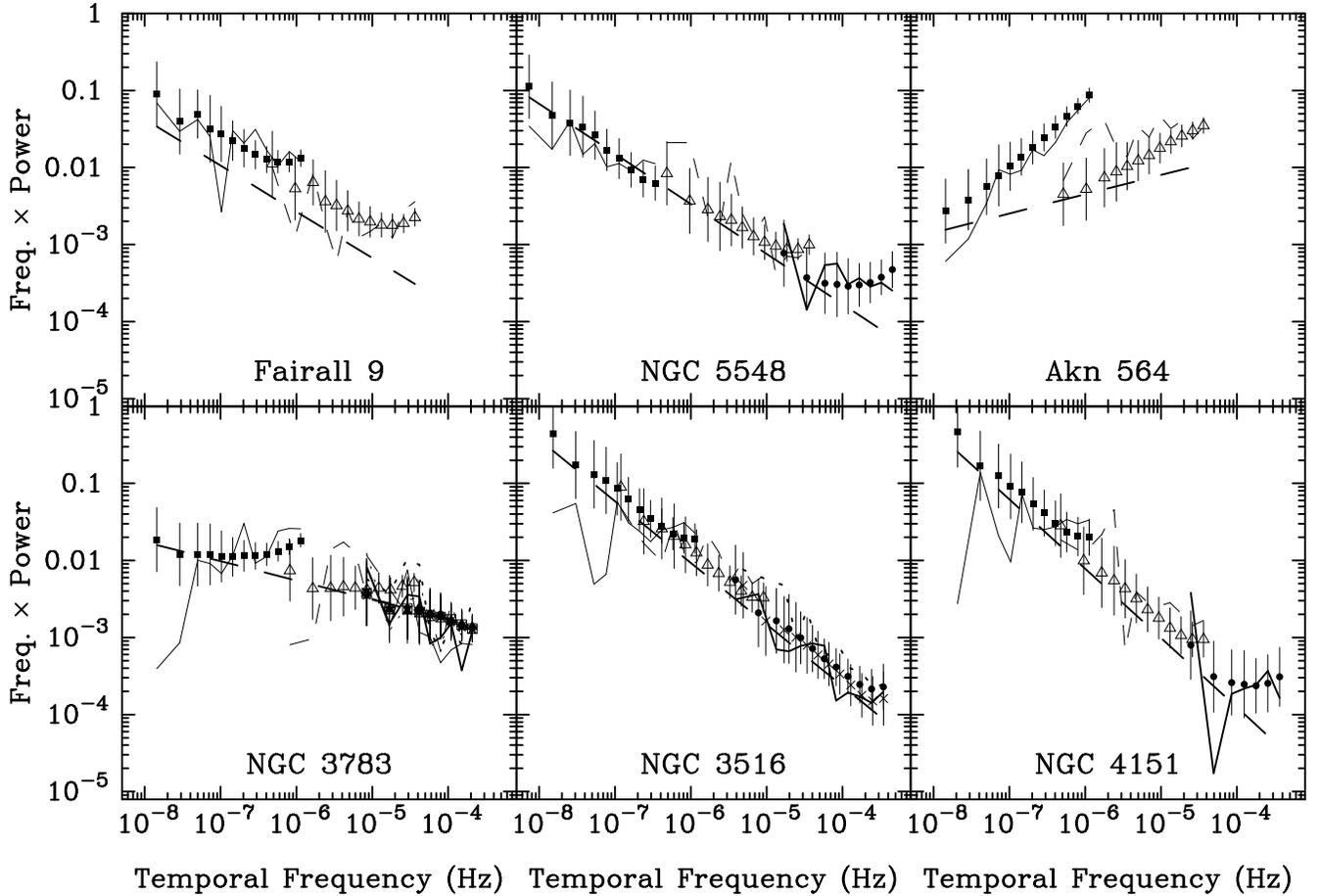}
\caption{Comparison of the best-fitting unbroken power law model
against the observed PSDs, plotted in $f~\times~P_f$ space.
The observed PSDs points are denoted by solid, dashed, or dotted
lines as described in Figure~5.
The best-fitting average distorted PSD model is denoted by symbols as
follows: filled squares denote PSD derived from long-term monitoring;
open triangles, medium-term monitoring; filled circles,
the short-term monitoring for NGC~5548, NGC~4151, and the 1997~long-look
of NGC~3516; crosses, the 1998~long-look
of NGC~3516. The five NGC~3783 {\it Chandra} PSDs, from
first to fifth, are marked by 
filled circles, crosses, open squares, filled stars, and filled triangles.
The bold dashed line represents the best-fitting unfolded PSD
model shape, excluding the distortion effects and Poisson noise,
which are evident in the observed and simulated PSD points.
Note, for instance, the large quantity of aliasing present in
the long-term PSD segment of Akn~564, due to the very high levels
of short-term variability in that object.}
\end{figure}

\begin{figure}[hb]
\epsscale{0.95}
\plotone{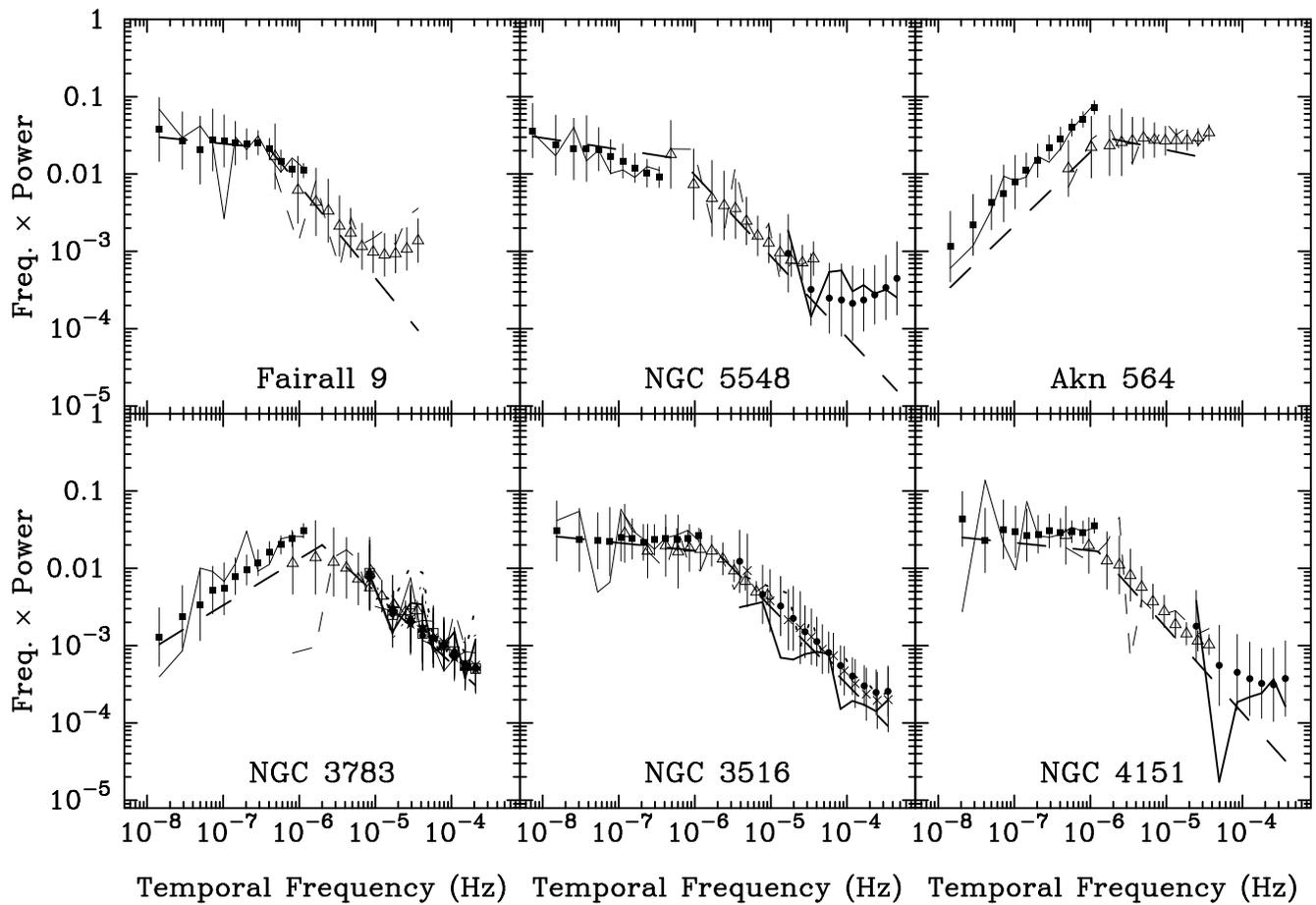}
\caption{Comparison of the best-fitting singly-broken power law model 
fits against the observed PSDs, plotted in $f~\times~P_f$ space. 
All lines and symbols
are the same as in
Figure~6.}
\end{figure}

\begin{figure}[hb]
\epsscale{0.95}
\plotone{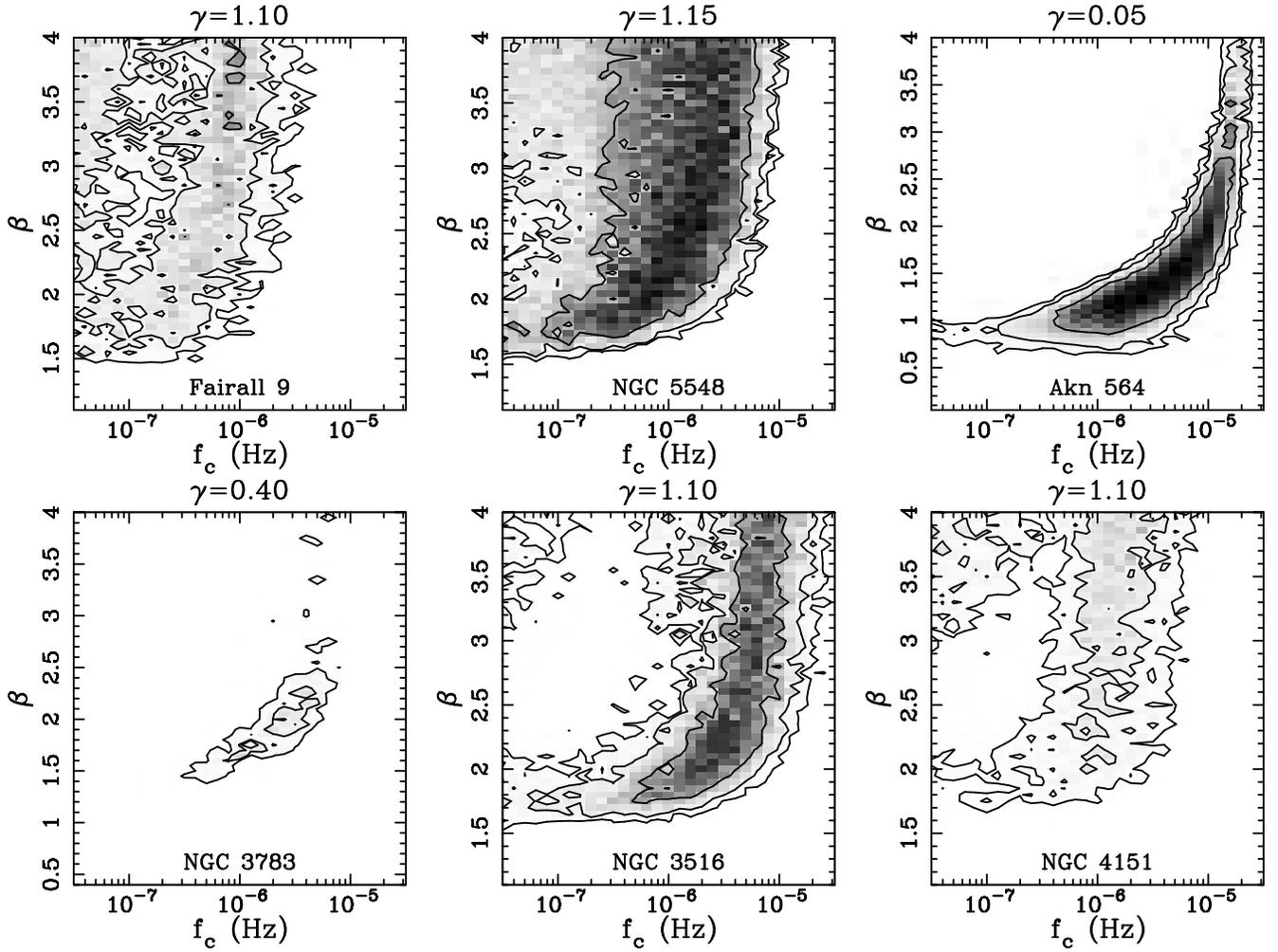}
\caption{Contour plots showing the errors on the best-fitting
singly-broken PSD model shape. For each target, a slice through
the three-dimensional fit parameter space at the best-fitting
low-frequency slope is shown. Solid lines
indicate the
99$\%$, 95$\%$, and 68$\%$ rejection probability levels.
Lighter shading denotes a high rejection probability;
darker shading denotes a low rejection probability.}
\end{figure}


\begin{figure}[hb]
\epsscale{0.94}
\plotone{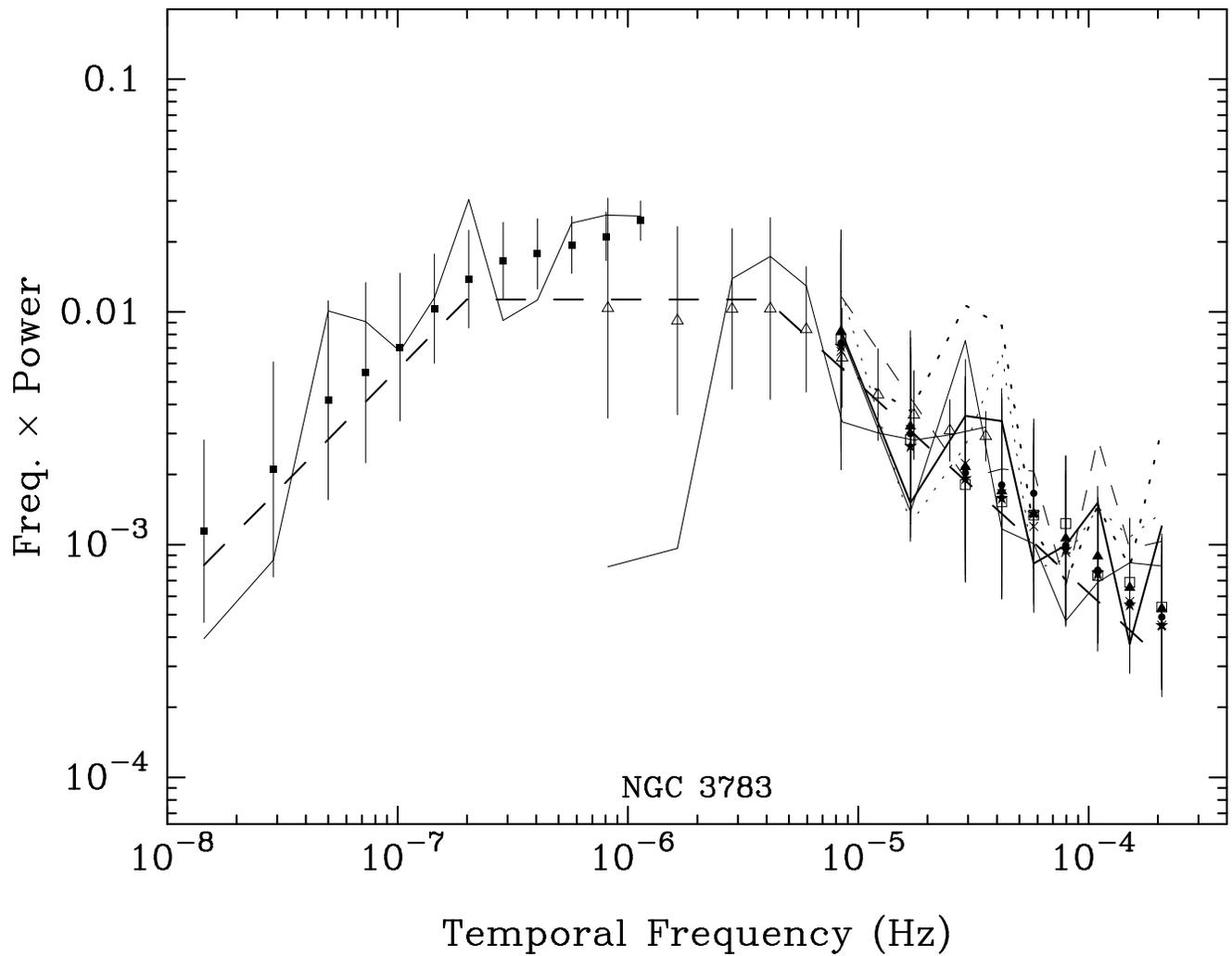}
\caption{Comparison of the best-fitting doubly-broken power law model fit
to the observed NGC~3783 PSD, plotted in $f~\times~P_f$ space. 
All lines and
symbols are the same as in
Figure~6.}
\end{figure}

\begin{figure}[hb]
\epsscale{0.83}
\plotone{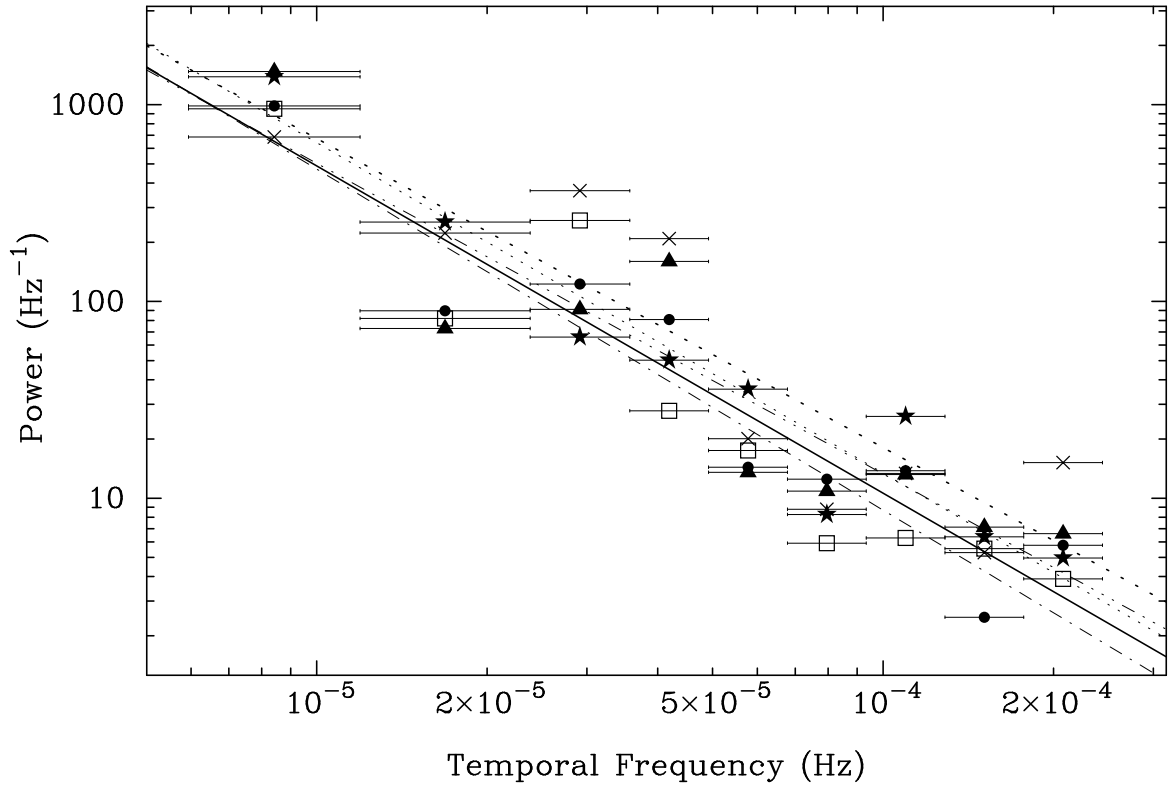}
\caption{ The five NGC~3783 \chandra\ short-term
binned PSDs. From the first-observed to last observed, the PSDs
are marked by filled circles, crosses, open squares, filled stars, and 
filled triangles.
The constant level of power due to Poisson noise has not been subtracted,
though the Poisson noise levels are very similar for each observation.
PSD errors are omitted here; they are determined later for a given
PSD model shape.
The best-fit lines (from first- to last-observed, respresented
by bold solid, 
bold dotted, solid, dashed, and dotted lines
respectively)
were calculated assuming equal weighting to all points,
but they are meant only to guide the eye, 
and are not an accurate representation of the 
intrinsic, underlying PSD slope
due to the presence of red-noise leak
and power due to Poisson noise.}

\end{figure}
\begin{figure}[hb]
\epsscale{0.95}
\plotone{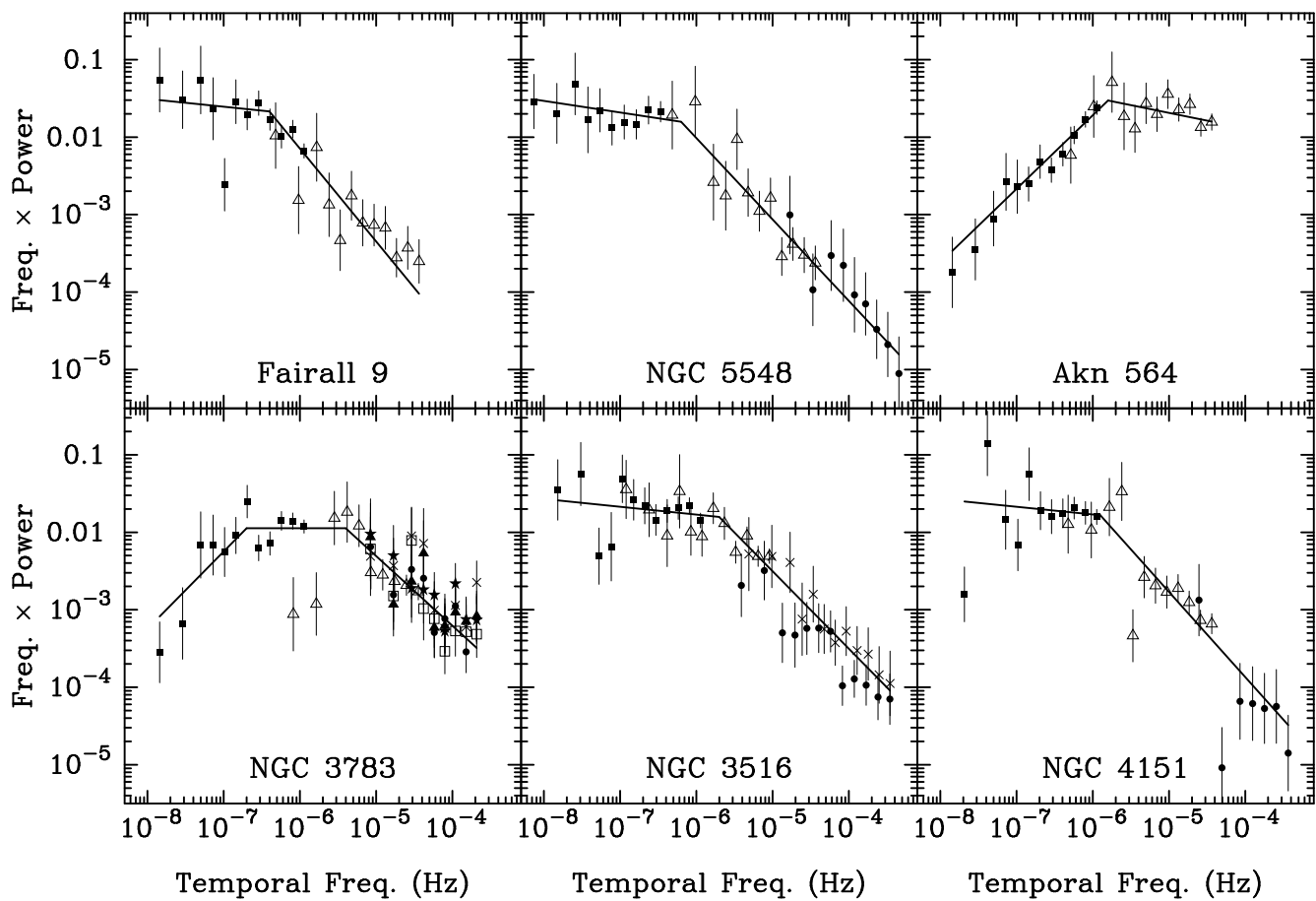}
\caption{The best-fitting model fits (doubly-broken power law model
for NGC~3783, singly-broken power law model for all other 
targets) are shown with the effects of Poisson noise, aliasing 
and red noise leak subtracted off. The solid line indicates the 
underlying, intrinsic PSD model shape. 
Symbols represent the differences between the average 
distorted model and the observed PSD data, plotted relative to the 
underlying PSD model shape.}
\end{figure}

\begin{figure}[hb]
\epsscale{0.85}
\plotone{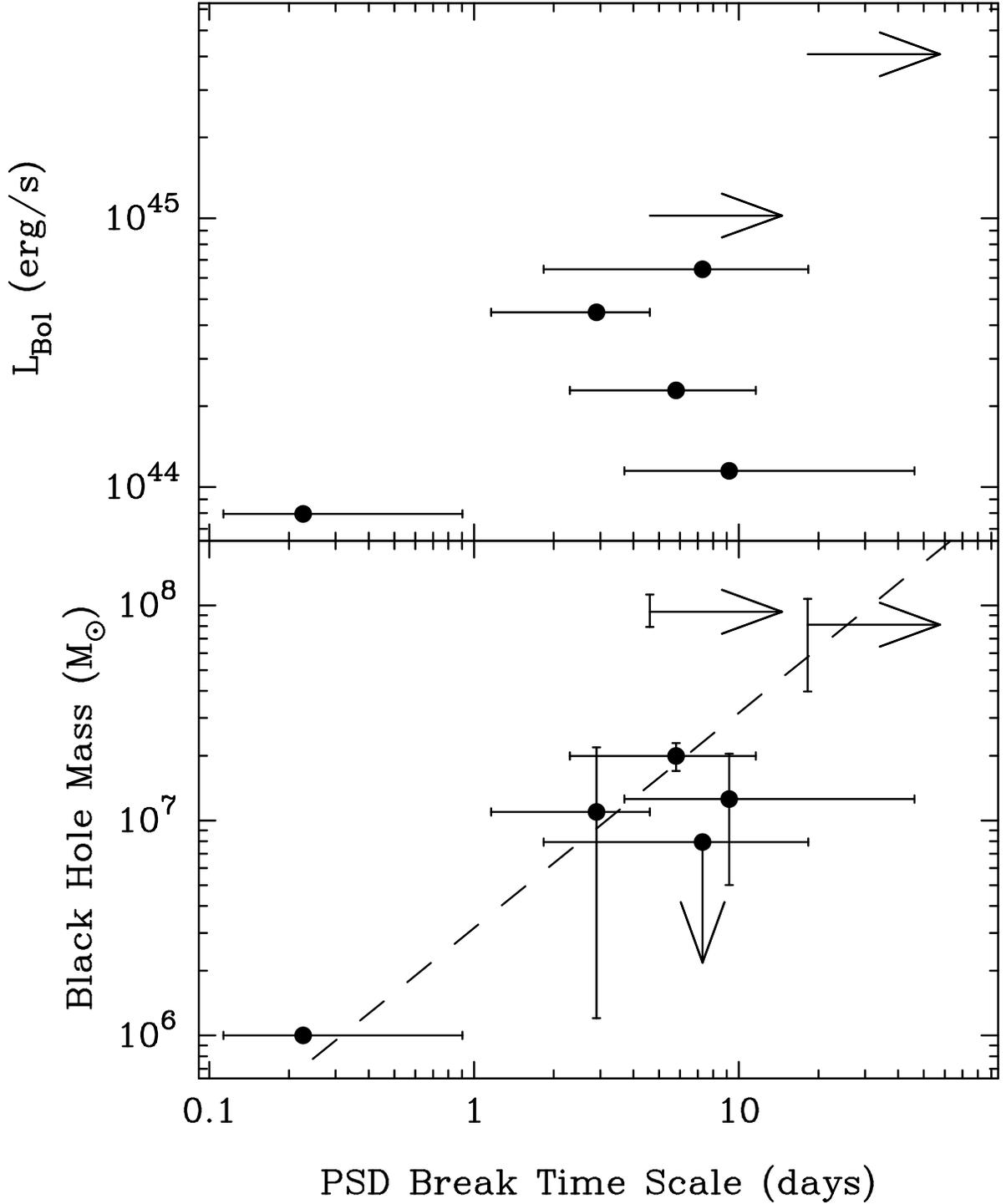}
\caption{Plot of bolometric luminosity and
estimated black hole mass plotted against
characterstic variability time scale (reciprocal 
of the PSD break frequency) from the singly-broken
power law model fits (for NGC~3783, the high-frequency break
from the doubly-broken model fit is used; lower limits to time scale
are shown for Fairall~9 and NGC~5548). The time scale for
MCG--6-30-15 is from UMP02.
Masses are reverberation mapping estimates 
from Kaspi \et\ (2000) except for NGC~3516 
(from Wanders \& Horne 1994) and
Akn~564 (rough reverberation estimate by Collier \et\ 2001).  
MCG--6-30-15 does not have a reverberation mass estimate;
see text for details.
The dashed line denotes the linear mass--time scale relation
$T_{\rm (days)} = M_{\rm BH}/10^{6.5} \Msun$.
}
\end{figure}

\begin{figure}[hb]
\epsscale{0.85}
\plotone{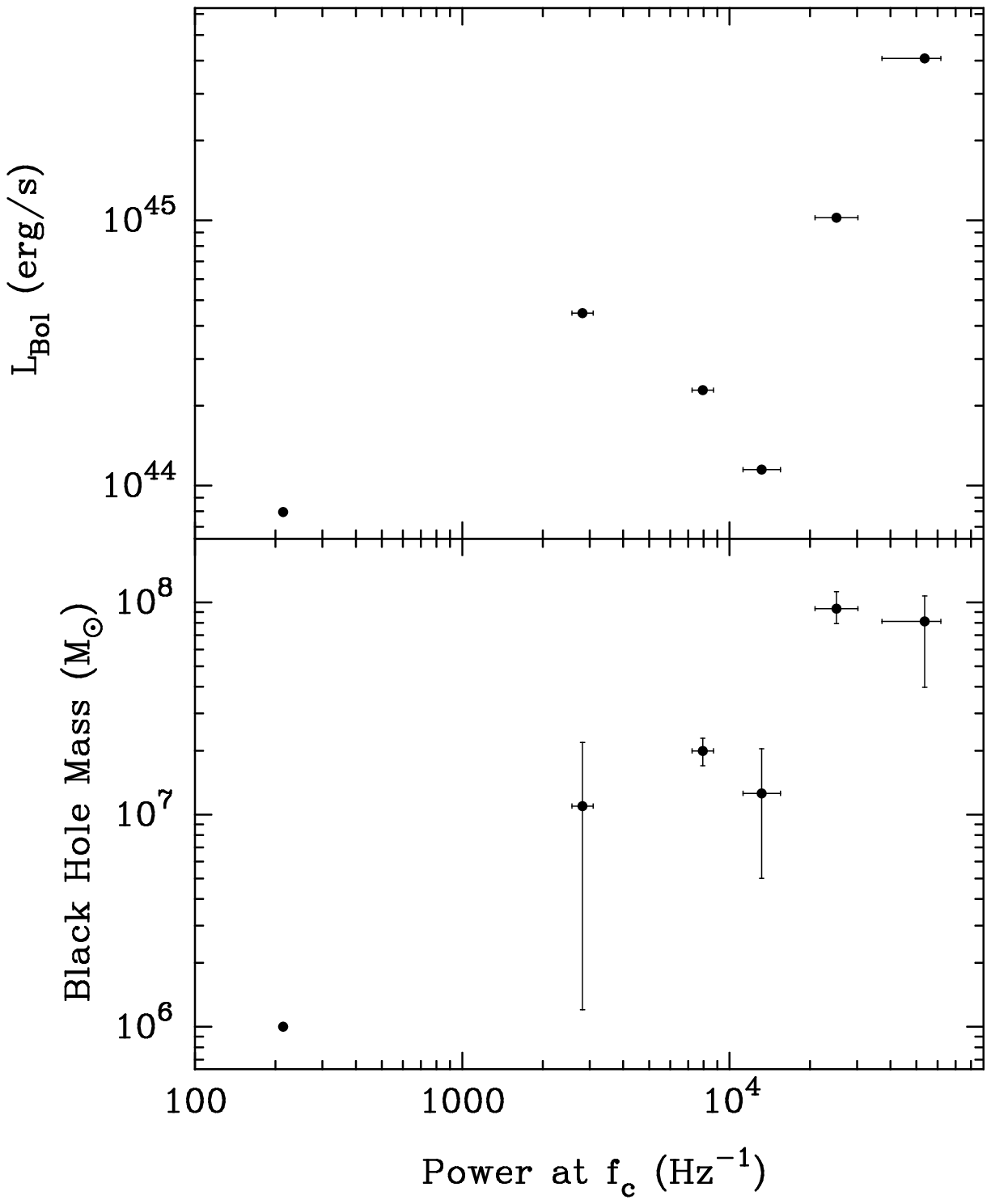}
\caption{Plot of bolometric luminosity and
estimated black hole mass plotted against the amplitude
of the best-fitting underlying PSD
at the break frequency. The model fit and amplitude
for MCG--6-30-15 is from UMP02.}
\end{figure}


\begin{references}

\reference{A88} Abramowicz, M.\ 1988, Advances in Space Research, 8, 151
\reference{B90} Belloni, T.\ \& Hasinger, G.\ 1990, A\&A, 230, 103
\reference{B81} Brillinger, D.\ 1981 ``Time Series: Data Analysis and
Theory,'' 2nd Edition (Holden-Day Publishing)
\reference{Ch01} Churazov, E., Gilfanov, M., \& Revnivstev, M.\ 2001 MNRAS, 321, 759
\reference{Col01} Collier, S., \et\ 2001, ApJ, 561, 146
\reference{D92} Done, C., Madejski, G., Mushotzky, R., Turner, T.\ J., Koyama, K., Kunieda, H.\ 1992, ApJ, 200, 138
\reference{E99} Edelson, R.\ \& Nandra, K.\ 1999, ApJ, 514, 682
\reference{E02} Edelson, R., Turner, J., Pounds, K., Vaughan, S., Markowitz, A., Marshall, H., Dobbie, P., 
Warwick, R.\ 2002, ApJ, 568, 610
\reference{Fn95} Fabian, A. C., Nandra, K., Reynolds, C. S.,
 Brandt, W. N., Otani, C., Tanaka, Y.,
 Inoue, H., Iwasawa, K.\ 1995, MNRAS, 277, L11
\reference{G93} Green, A., McHardy, I.\ \& Lehto, H.\ 1993, MNRAS, 265, 664
\reference{Ha98} Hayashida, K., Miyamoto, S., Kitamoto, S., \& Negoro, H.\ 1998, ApJ, 500, 642
\reference{H95} Herrero, A., Kudritzki, R., Gabler, R., Vilchez, J., \& Gabler, A.\ 1995, A\&A, 297, 556
\reference{K00} Kaspi, S., Smith, P., Netzer, H., Maoz, D., Jannuzi, B., \& Giveon, U.\ 2000, ApJ, 533, 631
\reference{K02} Kaspi, S.\ \et\ 2002, ApJ, 574, 643
\reference{L87} Lawrence, A., Watson, M., Pounds, K. \& Elvis, M.\ 1987, Nature, 325, 694
\reference{LP93} Lawrence, A.\ \& Papadakis, I.\ 1993, ApJL, 414, L85
\reference{L99} Leighly, K.\ 1999, ApJS, 125, 317
\reference{L97} Lyubarski, Y.\ 1997, MNRAS, 292, 679
\reference{M92} Maraschi, L., Molendi, S.\ \& Stella, L.\ 1992,
        MNRAS, 225, 27
\reference{Mk94} Markert, T., Canizares, C., Dewey, D., McGuirk, M., Pak, C \& Schattenburg, M.\ 1994, Proc.\ SPIE, 2280, 168
\reference{ME01} Markowitz, A.\  \& Edelson, R.\ 2001, ApJ, 547, 684
\reference{MC87} McHardy, I.\ \& Czerny, B.\ 1987, Nature, 325, 696
\reference{M88} McHardy, I.\ 1988., Memorie della Societa Astronomica Italiana, 59, 239
\reference{N97} Nandra, K., George, I., Mushotzky, R.F., Turner, T.J.,
        \& Yaqoob, T.\ 1997, ApJ, 476, 70
\reference{NP01} Nandra, K.\ \& Papadakis, I.\ 2001, ApJ, 554, 710
\reference{NY94} Narayan, R.\ \&  Yi, I.\ 1994, ApJL, 428, L13
\reference{N98} Nousek, J. \et\ 1998, Proc.\ SPIE, 3444, 225
\reference{N99a} Nowak, M., Vaughan, B., Wilms, J., Dove, J., \& Begelman, M.\ 1999a, ApJ, 510, 874
\reference{N99b} Nowak, M., Wilms, J., \& Dove, J.\ 1999b, ApJ, 517, 355
\reference{O75} Oppenheim, A.\ \& Shafer, R.\ 1975, ``Digital Signal
Processing,'' (Prentice-Hall Publishing)
\reference{PR88} Padovani, P.\ \& Rafanelli, P.\ 1988, A\&A, 205, 53
\reference{PL93a} Papadakis, I.\ \& Lawrence, A.\ 1993a, MNRAS, 261, 612
\reference{PL93b} Papadakis, I.\ \& Lawrence, A.\ 1993b, Nature, 361, 233
\reference{PM95} Papadakis, I.\ \& McHardy, I.\ 1995, MNRAS, 273, 923
\reference{PNK01} Papadakis, I., Nandra, K., \& Kazanas, D.\ 2001, ApJL, 554, L133
\reference{P02} Papadakis, I., Brinkmann, W., Negoro, H., \& Gliozzi, M.\ 2002, A\&AL, 382, L1
\reference{PW99} Peterson, B.\ \& Wandel, A.\ 1999, ApJL, 521, L95 
\reference{Pc82} Piccinotti, G. et al.\ 1982, ApJ, 253, 485
\reference{P80} Pineda, F. J., Delvaille, J. P., Schnopper, H. W., Grindlay, J. E. 1980, ApJ, 237, 414
\reference{P2} Plucinksy, P. \et\ 2002, "Astronomical Telescopes and Instrumentation 2002" (SPIE Conference Proceedings), eds. J.E.\ Truemper
and H.D.\ Tananbaum;  astro-ph/0209161
\reference{P95} Pounds, K., Done, C.\ \& Osborne, J.\ 1995, \mnras, 277, L5
\reference{P01} Pounds, K., Edelson, R., Markowitz, A., \&  Vaughan, S.\ 2001, ApJL, 550, L15
\reference{PF99} Poutanen, J. \& Fabian, A.\ 1999, MNRAS, 306, L31
\reference{Pr81} Priestley, M.\ 1981, ``Spectral Analysis and Time Series,'' (London: Academic Press Ltd.)
\reference{R97} Reynolds, C.,\  Ward, M., Fabian, A., \& Celotti, A.\ 1997, MNRAS, 291, 403
\reference{R00} Reynolds, C.\ 2000, ApJ, 533, 811
\reference{njs02} Schurch, N. \& Warwick, R.\ 2002, MNRAS, 334, 811
\reference{SS73} Shakura, N. I., \& Sunyaev, R.A.\ 1973, A\&A, 24, 337
\reference{S01} Struder, L.\ \et\ 2001, A\&AL, 365, L27
\reference{Sw98} Swank, J.\ 1998, in Nuclear Phys. B (Proc. Suppl.): The Active X-ray Sky: Results From BeppoSAX and Rossi-XTE, Rome, Italy, 1997 October 21-24, eds. L. Scarsi, H. Bradt, P. Giommi, \& F. Fiore, Nucl.\ Phys. B Suppl.\ Proc.\ (The Netherlands: Elsevier Science B.V.), 69, 12 
\reference{SR00} Sunyaev, R.\ \& Revnivtsev, M.\ 2000, A\&A, 358, 617 
\reference{Tn95} Tanaka, Y., Nandra, K., Fabian, A. C.,
 Inoue, H., Otani, C., Dotani, T., Hayashida, K.,
 Iwasawa, K., Kii, T., Kunieda, H., Makino, F., \&
 Matsuoka, M.\ 1995, Nature, 375, 659
\reference{TK95} Timmer, J.\ \& K\"{o}nig, M.\ 1995, A\&A 300, 707
\reference{T88} Treves, A., Maraschi, L., Abramowicz, M.\ 
        1988, PASP, 100, 427
\reference{T01} Turner, M.\ J.\ L.\ \et\ 2001,  A\&AL, 365, L18 
\reference{T99} Turner, T. J., George, I. M., Nandra, K., \& Turcan, D.\ 1999, ApJ, 524, 667
\reference{T02} Turner, T.\ J., \et\ 2002, ApJ, 568, 120
\reference{UM01} Uttley, P.\ \& McHardy, I.\ 2001, MNRAS, 323, L26
\reference{UMP02} Uttley, P., McHardy, I.\ \& Papadakis, I.\ 2002, MNRAS, 332, 231
\reference{v97} van der Klis, M.\ 1997, in G.\ J.\ Babu, E.\ D.\ Feigelson, eds., 'Statistical Challenges in Astronomy II,' Springer-Verlag (New York), p.\ 321
\reference{V99} Vaughan, S., Pounds, K., Reeves, J., Warwick, R., Edelson, R.\ 1999, MNRAS, 304, L34 
\reference{W02} Wandel, A. 2002, ApJ, 565, 762
\reference{WH94} Wanders, I.\ \& Horne, K.\ 1994, A\&A, 289, 76
\reference{WE87} Wilkes, B. \& Elvis, M.\ 1987, ApJ, 323, 243
\reference{Wi99} Wilms, J., Nowak, M.,
 Dove, J., Fender, R.,
 di Matteo, T.\ 1999, ApJ, 522, 460
\end{references}
\end{document}